\documentclass[fleqn,usenatbib]{mnras}

\usepackage{newtxtext,newtxmath}
\usepackage{ulem}
\usepackage[T1]{fontenc}

\DeclareRobustCommand{\VAN}[3]{#2}
\let\VANthebibliography\thebibliography
\def\thebibliography{\DeclareRobustCommand{\VAN}[3]{##3}\VANthebibliography}

\usepackage{graphicx}	% Including figure files
\usepackage{amsmath}	% Advanced maths commands
\usepackage{csvsimple}
\usepackage{tabularx}
\usepackage{geometry}

\usepackage{longtable}
\usepackage{booktabs}
\usepackage{multirow}
\usepackage{comment}
\usepackage{tikz,xcolor,hyperref}
\definecolor{lime}{HTML}{A6CE39}
\DeclareRobustCommand{\orcidicon}{%
	\begin{tikzpicture}
	\draw[lime, fill=lime] (0,0) 
	circle [radius=0.16] 
	node[white] {{\fontfamily{qag}\selectfont \tiny ID}};
	\draw[white, fill=white] (-0.0625,0.095) 
	circle [radius=0.007];
	\end{tikzpicture}
	\hspace{-2mm}
}
\foreach \x in {A, ..., Z}{%
	\expandafter\xdef\csname orcid\x\endcsname{\noexpand\href{https://orcid.org/\csname orcidauthor\x\endcsname}{\noexpand\orcidicon}}
}

\title[Disc-jet coupling in blazar]{Probing the disc-jet coupling in S4 0954+65, PKS 0903-57, $\&$ 4C +01.02 with $\gamma$-rays}

\author[A. Sharma et al.]{
Ajay Sharma\orcidA,$^{1}$\thanks{ajjjkhoj@gmail.com }
Sushanth Reddy Kamaram\orcidB,$^{2}$\thanks{ksushanthreddy1999@gmail.com}
Raj Prince\orcidC,$^{3}$\thanks{raj@cft.edu.pl}
Rukaiya Khatoon,$^{4}$\thanks{rukaiyakhatoon12@gmail.com}
Debanjan Bose\orcidE$^{5}$\thanks{debaice@gmail.com}
\\
$^{1}$ S. N. Bose National Centre for Basic Sciences, Block JD, Sector III, Salt Lake, Kolkata 700 106, India\\
$^{2}$ Indian Institute of Technology Kharagpur, 721302, India\\
$^{3}$ Center for Theoretical Physics, Polish Academy of Sciences, Al. Lotnikow, 32/46, Warsaw, Poland\\
$^{4}$ Center for Space Research, North-West University, Potchefstroom, 2520, South Africa\\
$^{5}$ School of Astrophysics, Presidency University, Kolkata 700073, India
}

\date{Accepted XXX. Received YYY; in original form ZZZ}

\pubyear{2015}

\begin{document}
\label{firstpage}
\pagerange{\pageref{firstpage}--\pageref{lastpage}}
\maketitle

\begin{abstract}
We present a comprehensive variability study on three blazars, S4 0954+65, PKS 0903-57, and 4C +01.02 covering a mass range of log(M/M$_{\odot}$) = 8--9, by using $\sim$15 years-long $\gamma$-ray light curves from \textit{Fermi}-LAT. The variability level is characterized by the fractional variability amplitude which is higher for $\gamma$-rays compared to optical/UV and X-rays emissions. A power spectral density (PSD) study and damped random walk (DRW) modeling are done to probe the characteristic timescale. The PSD is fitted with a single power-law (PL) and bending power-law models and the corresponding success fraction was estimated. %\sout{The PSD of PKS 0903-57 shows a break at 250 days. A break at 150 days is also observed in S4 0954+65 but both models have similar success fractions.} %\textbf{PSD and DRW modeling stand as independent methodologies for characterizing the break feature in the power spectral density of $\gamma$-ray light curves. 
In the case of PKS 0903-57, We observed a break in the $\gamma$-ray PSD at 256 days which is comparable to the viscous timescale in the accretion disc suggesting a possible disk-jet coupling.
%the break feature obtained through PSD modeling is consistent with the observed break nature in PSD constructed from DRW modeling outcomes,} and the break timescale is also consistent with the viscous timescale in the accretion disc. 
%%%Our results show that the PSD break time scale in PKS 0903-57 is consistent with the viscous time scale in the accretion disc.%%%
%results are consistent with \textit{flicker noise} obtained from single PL - indicative of the underlying nature of variability due to persistent long-memory processes and also obtained breaks in frequency for S4 0954+65 and PKS 0903-57 at 0.006 ($\mathrm{day^{-1}}$) and 0.004 ($\mathrm{day^{-1}}$) respectively. 
%To further validate the observed characteristic variability timescales of blazars, we extended the confirmation by generating accurate PSDs through the application of the stochastic (DRW) model on the $\mathrm{\gamma}$-ray LC. 
%\sout{The PSD derived from the DRW modeling shows a similar break as obtained in the independent PSD analysis.} 
The non-thermal damping timescale from the DRW modeling is compared with the thermal damping timescale for AGNs including our three sources. Our sources lie on the best-fit of the $\mathrm{\tau^{rest}_{damping}} - M_{BH}$ plot derived for AGN suggesting a possible accretion disc-jet connection. If the jet's variability is linked to the disc's variability, we expect a log-normal flux distribution, often connected to the accretion disc's multiplicative processes. Our study observed a double log-normal flux distribution, possibly linked to long and short-term variability from the accretion disk and the jet. In summary, PSD and DRW modeling results for these three sources combined with blazars and AGNs studied in literature favor a disc-jet coupling scenario. However, more such studies are needed to refine this understanding.  
\end{abstract}


%%%%%%%%%%%%%%%%%%%%%%%%%%%%%%%%%%%%%%%%%%%%%%%%%%

%%%%%%%%%%%%%%%%% BODY OF PAPER %%%%%%%%%%%%%%%%%%

\section{Introduction}

% This is a simple template for authors to write new MNRAS papers.
% See \texttt{mnras\_sample.tex} for a more complex example, and \texttt{mnras\_guide.tex}
% for a full user guide.

% All papers should start with an Introduction section, which sets the work
% in context, cites relevant earlier studies in the field by \citet{Fournier1901},
% and describes the problem the authors aim to solve \citep[e.g.][]{vanDijk1902}.
% Multiple citations can be joined in a simple way like \citet{deLaguarde1903, delaGuarde1904}.

Active galactic nuclei (AGN) are the most luminous sources in the universe, hosting supermassive black holes (SMBHs) at their centers. These extraordinary astronomical sources have been studied for more than half a century \citep{matthews1963optical}, and can be distinguished and identified through a range of factors, including their bolometric luminosity ($L \sim 10^{47} erg s^{-1}$), flux variabilities, and spectral energy distribution (SEDs), etc. The immense luminosity from AGN arises from the accretion of matter around the SMBHs. This process serves as the primary source of energy that powers the AGN. Flux variability involves non-periodic variations in flux, which occur on various timescales and exhibit different amplitudes. Only a small fraction of AGN show radio emissions and are called radio-loud AGN. Some of these AGN have relativistic jets pointed towards Earth and are known as Blazars \citep{urry1995unified}. These sources are known to exhibit extreme properties such as high-energy emission, and rapid and large amplitude flux variations in the entire accessible spectral range \citep{ulrich1997variability}. Blazars are also sources of non-thermal emissions, which undergo Doppler boosting due to their relativistic jets. %This Doppler boosting is manifested in their exceptional brightness and high variability in the vast spatial and temporal frequency domains.% \par %{ \textcolor{green}{The dominance of blazars in the extra-galactic $\gamma$-ray sky has been unequivocally established through two distinct experiments: the Compton Gamma Ray Observatory (CGRO) \citep{hartman1999third} and the Fermi Gamma Ray Space Telescope \citep{lat2019fermi}. Furthermore, these objects are likely sources to produce extra-solar neutrinos \citep{icecube2018neutrino, icecube2018multimessenger}}. \textcolor{cyan}{Blazars are further classified into two distinct subgroups: BL Lacertae (BL Lacs) and flat spectrum radio quasars (FSRQs) objects. Within the classification of blazars, Flat Spectrum Radio Quasars (FSRQs) display broad emission lines in their optical spectra. While, BL Lacs exhibit either featureless optical spectra or optical spectra containing weak emission lines (equivalent width $< 5 \mathring{A} $). Alternatively, \citep{ghisellini2011transition} proposed a robust distinction criterion between FSRQs and BL Lacs based on the luminosity of the Broad Line Region (BLR) relative to the Eddington luminosity. FSRQs possess a luminosity ratio with $\frac{L_{BLR}}{L_{Edd}}> 5 \times 10^{-5} $, indicating their status as beamed counterparts of the more luminous Fanaroff and Riley type $FR2$ \citep{fanaroff1974morphology} radio sources. While BL Lacs represent the less luminous beamed counterparts of $FR1$ radio sources. The broad-band spectral energy distribution (SED) of blazars typically exhibits two distinct spectral components. The first component, known as the low-energy segment, spans from radio to optical wavelengths (and sometimes extending upto X-rays for BL Lacs).} This peak is unambiguously associated with synchrotron emission arising from the acceleration of relativistic particles within the source. Whereas, the second component, known as the high-energy segment, is typically observed between ultraviolet (UV)/X-rays and $\gamma$-ray energies (upto GeV/TeV's). This peak is believed to originate from the process of inverse Compton scattering \citep{bottcher2019progress} of low energy photons by electrons. If synchrotron photons get up-scattered by same electrons then that process is called  synchrotron self-compton (SSC) \citep{mastichiadis2002models} process. Electrons can also up-scatter external photons (where the soft photons were produced by various regions of blazars, such as an accretion disk, broad-line region, and dusty torus) through external compton (EC) process within the jets. Blazars can be further classified into sub-groups based on the location of the peak of the synchrotron emission in their broad-band spectral energy distribution (SED) \citep{fan2016spectral}. If the first hump of the SED is peaked at $\nu_s < 10^{14} Hz$, they are referred to as low-synchrotron peaked (LSP) blazars. Intermediate synchrotron peaked (ISP) blazars are characterized by the peak of the synchrotron hump in between $10^{14} Hz \le \nu_s \le 10^{15} Hz$, while high-synchrotron peaked (HSP) blazars have their synchrotron hump peak at $\nu_s > 10^{15} Hz$. The previously discussed mechanisms for high-energy emission in blazars primarily pertain to the "leptonic" scenarios \citep{madejski2016gamma}. However, in the "Hadronic" scenario, an alternative explanation for the high-energy emission arises. In this scenario, the peak of the high-energy emission is attributed to protons that have been accelerated to $\simeq PeV-EeV$ energies. These energetic protons can generate $\gamma$-rays through either the synchrotron process or meson decay. Furthermore, the secondary particles produced through proton-proton interactions can also contribute to the observed high energy emission. The inclusion of the Hadronic scenario provides an additional perspective for understanding the diverse processes and sources of high-energy emissions in blazars \citep{bottcher2013leptonic}.}\par

%Blazars show flux variability across the electromagnetic spectrum, ranging from radio to very high-energy (VHE) $\gamma$-rays. 
Multi-frequency variability studies have emerged as a vital tool for gaining valuable insights into the physical conditions prevailing within the innermost regions of blazars. These studies provide a means to investigate various aspects of blazars, such as the dominant particle acceleration mechanism, energy dissipation processes, magnetic field geometry, and the composition of the jet. The flux variability is classified into sub-categories: long-term variability, occurring over timescales of decades/years to several months; short-term variability, manifesting within weeks to days; and intranight variability, which transpires within a single day \citep{ulrich1997variability}. The power spectral densities (PSDs) from blazar light curves often exhibit a distinctive pattern characterized by a single power-law shape, represented by a form $P(\nu_k) \propto \nu_{k}^{-\beta}$, where $\beta$ denotes the slope of the power-law function, while $\nu_k$ represents the temporal frequencies. This behavior indicates that the observed variability follows a colored noise-type stochastic process \citep{finke2014fourier, goyal2017multiwavelength}. Specifically, different values of $\beta$ offer insights into distinct types of stochastic processes. For instance, when $\beta \simeq$ 1, $\beta \simeq$ 2, and $\beta \gtrsim $ 3, we observe long memory/pink-noise, damped walk/red-noise, and black noise-type stochastic processes, respectively. While $\beta \sim$ 0 represents uncorrelated white-noise type stochastic process \citep{press1978flicker}. Fluctuations arising from the stochastic process follow certain probability distributions, which impart predictability to the PSDs. Therefore, the PSD of the light curves holds valuable information about the parameters of the stochastic process and the characteristic/relaxation timescales governing the variability in the system. By examining the slope, normalization, and breaks in the PSD, we can glean insights into various aspects, such as the timescales associated with particle cooling or escape \citep{kastendieck2011long, sobolewska2014stochastic, finke2014fourier, chen2016particle, kushwaha2017gamma, chatterjee2018possible, ryan2019characteristic, bhattacharyya2020blazar}. Significant efforts have been made to model the complex phenomena related to the multi-waveband variability/SED observed in the sources. These models encompass a diverse range of physical processes that occur either within the accretion disc or the jets. The proposed scenarios explore emission sites located in the accretion disc, which revolves around a supermassive black hole, magneto-hydrodynamic instabilities within the disc and jets, the propagation of shocks downstream in relativistic jets, as well as the relativistic effects resulting from the orientation of the jet. These models aim to capture the intricate mechanisms responsible for the observed variability, taking into account the interplay between the accretion disc and the jet and their influence on the emission across different wavebands \citep{camenzind1992lighthouse, wagner1995intraday, marscher2013turbulent}. However, the precise study of the underlying processes is still a subject of ongoing debate and investigation.\par
Extensive studies utilizing large datasets across various wavebands have been conducted to characterize the properties of PSDs over a wide range of timescales, spanning from days to years. At the radio frequencies, \citet{park2017long, max2014method} analyzed decades-long radio data and reported $\beta \sim$ 1-3. At optical frequencies, \citet{nilsson2018long, goyal2021optical} reported $\beta \sim$ 1-1.5. In the X-ray waveband, \citet{kataoka2001characteristic, isobe2014maxi, chatterjee2018possible} reported $\beta \sim$ 1-1.5. Studies in the high-energy $\gamma$-rays domain, \citep{abdo2010gamma, meyer2019characterizing, bhatta2020nature, tarnopolski2020comprehensive} reported $\beta \sim$ 1-2. In the very high-energy $\gamma$-ray domain, \citet{goyal2020blazar} reported $\beta \sim$ 1 for couple of blazars.\par
%\textcolor{red}{New paragraph added for DRW model}. 
Another approach involves fitting stochastic light curves using Gaussian processes in the time domain \citep{kelly2009variations, li2018new, macleod2010modeling}. This approach has gained considerable traction in characterizing the optical variability of AGNs. A notable representation of such processes is the Continuous-time Autoregressive Moving Average (CARMA) models developed by \citet{kelly2014flexible}. The damped random walk (DRW) model, also represented by CARMA(1,0), has demonstrated its efficacy in accurately characterizing the long-term variability of the AGN accretion discs \citep{kelly2009variations, kozlowski2009quantifying, zu2013quasar, rakshit2017optical, kasliwal2017extracting, lu2019supermassive, Burke_2021}. These stochastic processes have proven to be powerful methods for probing AGN variability with remarkable success. Recent advances have extended the utility of the CARMA model to the realm of $\gamma$-ray variabilities of AGN \citep{sobolewska2014stochastic, goyal2018stochastic, ryan2019characteristic, tarnopolski2020comprehensive, covino2020looking, Yang_2021, zhang2022characterizing}. This extension has demonstrated the ability to capture features of the flux variability and to provide a more accurate representation of the PSD. In addition to CARMA, \citet{foreman2017fast} developed another fast and flexible method for light curve modeling with a stochastic process, \texttt{celerite}. With access to almost 15 years of Fermi-LAT data, it provides an opportunity to comprehensively study the $\gamma$-ray light curves of AGN. This endeavor allows us to thoroughly investigate and reveal the inherent stochastic properties through well-established methods. \par
The main objective of this study is to comprehensively examine and understand the disk-jet coupling in blazar.
%statistical properties of multifrequency variability in blazars.
To achieve this, we conduct an extensive investigation utilizing 15 years long data from the Fermi-LAT $\&$ Swift-XRT/UVOT observations. We focus on three selected blazars S4 0954+65, PKS 0903-57, and 4C +01.02 of different types (BL Lacertae [BL Lac], Blazar Candidates of Uncertain class [BCU], and Flat Spectrum Radio Quasar [FSRQ]) and covering a mass range from 10$^8$ to 3$\times$10$^9$ M$_{\odot}$. In section 2, the data analysis methods for Fermi-LAT and Swift-XRT/UVOT are outlined. In section 3, A variety of analysis approaches have been used to comprehensively investigate the broadband light curves. These methods include fractional variability, flux distribution, PSD, and DRW modeling analyses, and the obtained results are detailed in corresponding sections. In section 4, we discuss the obtained results in detail. Finally, A comprehensive conclusion is presented in Section 5.

\section{MULTI-WAVELENGTH OBSERVATIONS AND DATA ANALYSIS}

In this investigation, a sample of blazars (S4 0954+65, PKS 0903-57, and 4C +01.02) has been selected for a comprehensive study of multi-waveband variability within the time span of MJD 54683 - 59970. These sources are positioned at (RA: 149.69, Dec: 65.56) for S4 0954+65 with redshift, z $\sim$ 0.367 \citep{lawrence1986new,lawrence1996optical,stickel1993complete}, (RA: 136.22, Dec: -57.58) for PKS 0903-57 with z $\sim$ 0.695 \citep{thompson1990spectroscopy}, and (RA: 17.16, Dec: 1.58) for 4C +01.02 with z $\sim$ 2.099 \citep{hewett1995large}. %In this study, we utilized data extracted from within $10^\circ$ region of interest(ROI). this information is irrelevant here}
The BH mass of S4 0954+65 is $\sim 3.3 \times 10^8 M_{\odot} $, estimated from the width of the H$_{\alpha}$ line \citep{fan2004black}. For 4C +01.02, a BH mass of $\sim 3 \times 10^9 M_{\odot}$ was found through SED and spectropolarimetry modeling \citep{schutte2022modeling}. For PKS 0903-57, we did not find an estimated BH mass value. However, we have adopted the most probable BH mass value between $3-6 \times 10^8 M_{\odot}$ from the distribution of BH mass of all radio-loud AGN from \citet{Ghisellini2015}.

\begin{figure*}
\centering
\includegraphics[width=0.90\textwidth]{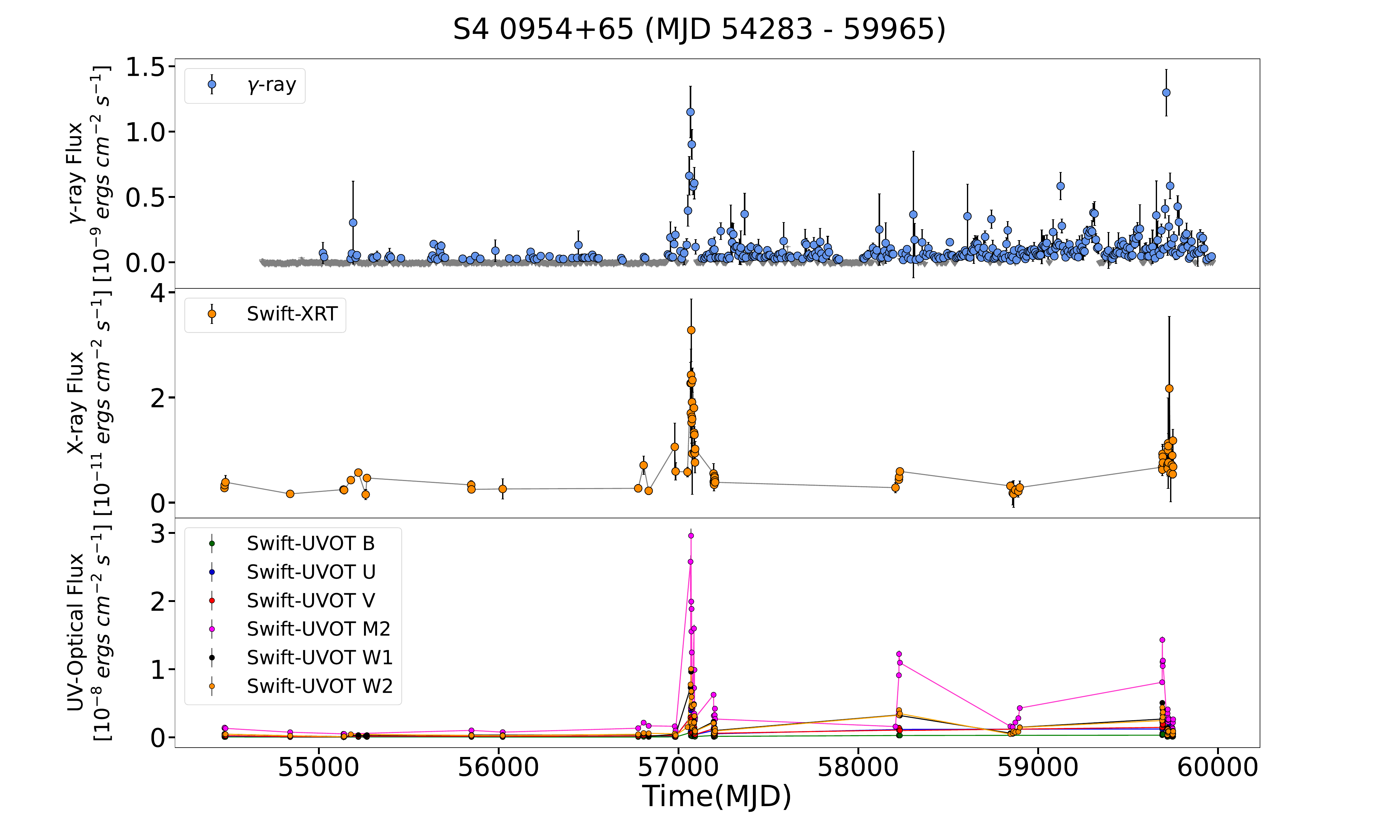}
\caption{The multi-wavelength light curve of blazar S4 0954+65 in the time domain MJD 54283 - 59965. The top panel displays the weekly binned $\gamma$-ray light curve obtained from Fermi-LAT data.  The grey points represent the upper limits. The middle panel presents the X-ray light curve generated using all available observation IDs from July 2007 to January 2022. Finally, the bottom panel shows the light curves for all Swift-UVOT bands (V, B, U, UVM2, UVW1, UVW2), which were generated from the same observations as the Swift-XRT data.}
\label{fig:S4_lc}
\end{figure*}
%{S40954+65_gammaergs+XRT+UVOT_LC.pdf}
\begin{figure*}
\centering
\includegraphics[width=0.90\textwidth]{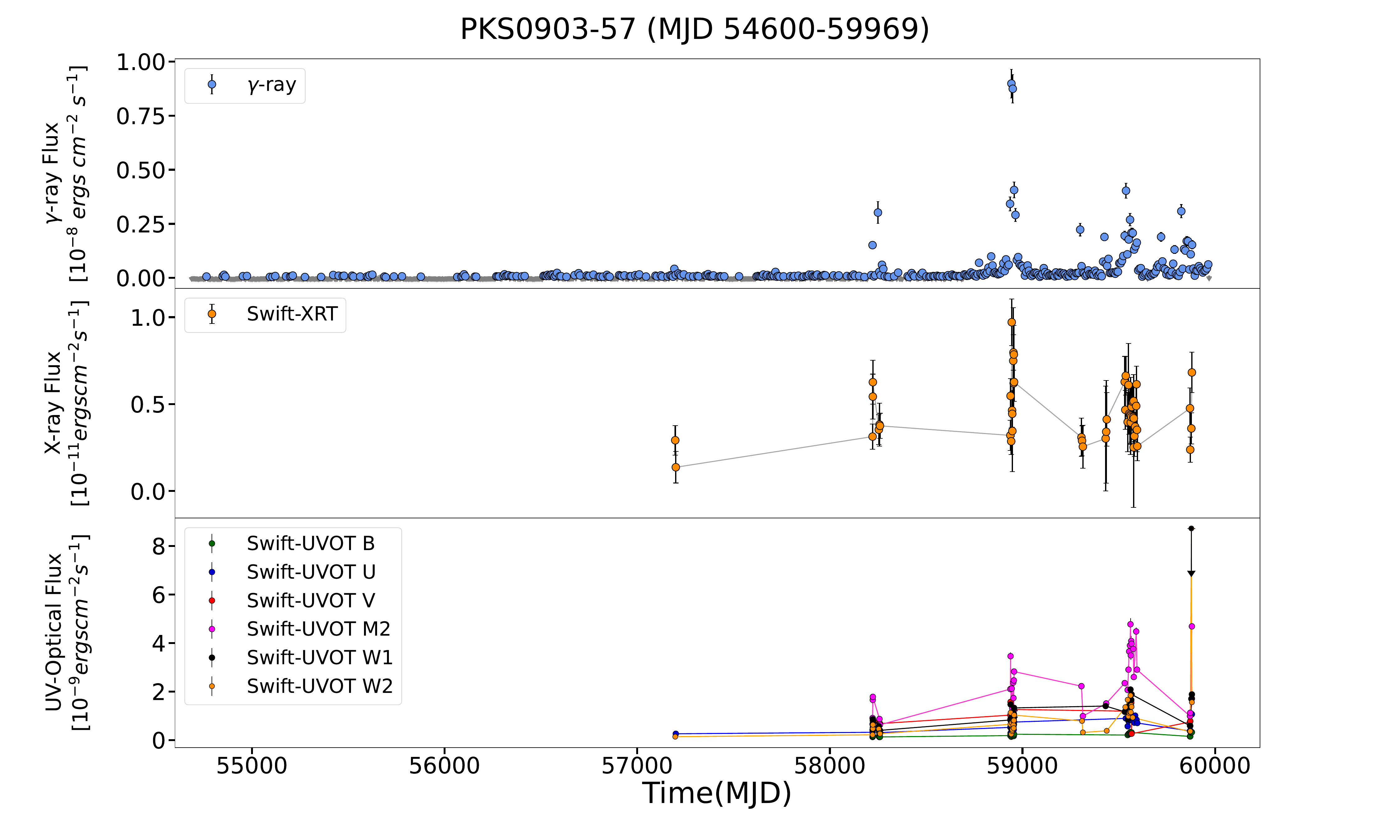}
\caption{Shows the multi-wavelength light curve of the blazar PKS 0903-57 in the time domain MJD 54600 - 59969. The top panel displays the weekly binned $\gamma$-ray light curve obtained from Fermi-LAT data. The grey points represent the upper limits. The middle panel presents the X-ray light curve generated using all available observation IDs from May 2008 to January 2023. Finally, the bottom panel shows the light curves for all Swift-UVOT bands (V, B, U, UVM2, UVW1, UVW2), which were generated from the same observations as the Swift-XRT data.}
\label{fig:pks09_lc}
\end{figure*}

\begin{figure*}
\centering
\includegraphics[width=0.95\textwidth]{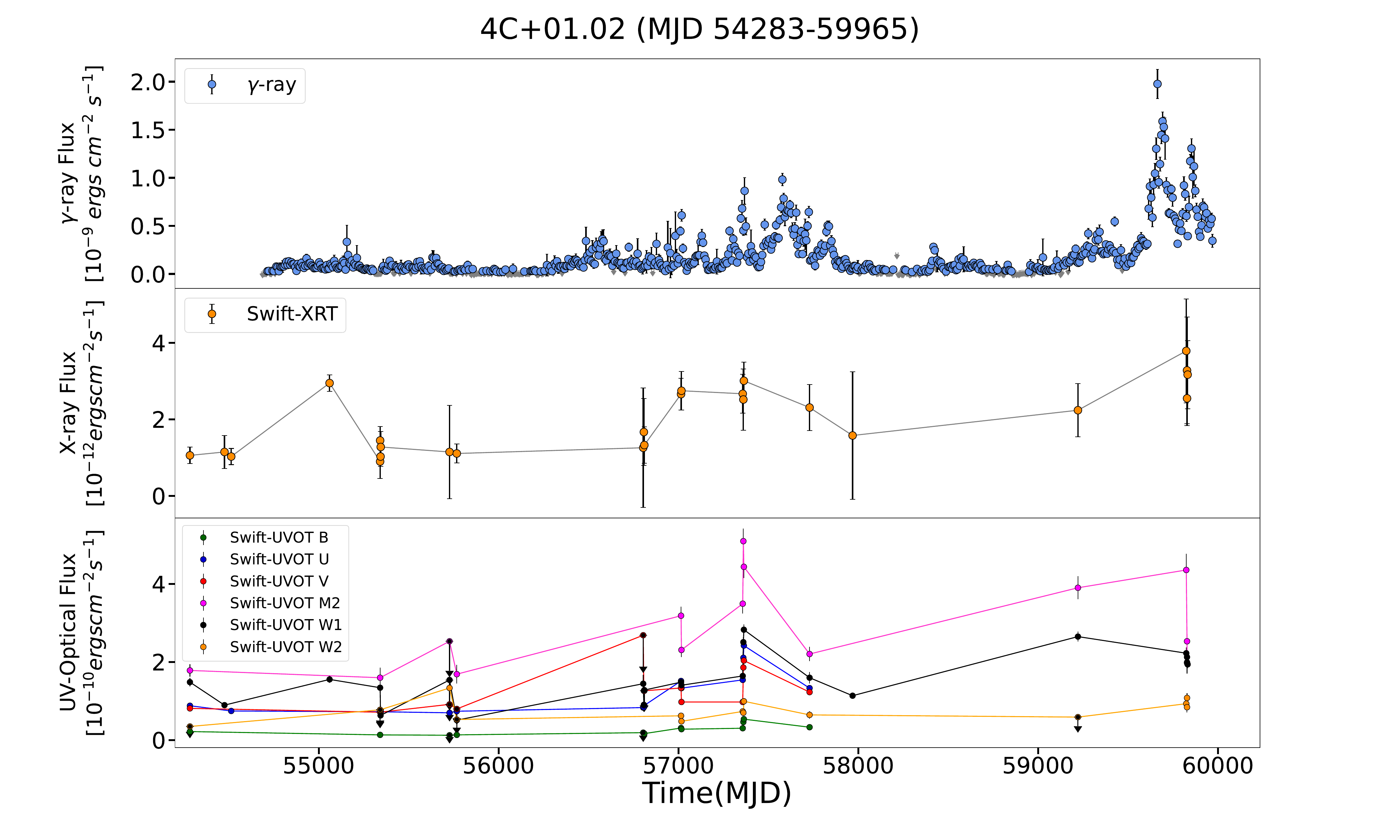}
\caption{Shows the multi-wavelength light curve of 4C +01.02 in the time domain MJD 54283 - 59965. The top panel displays the weekly binned $\gamma$-ray light curve obtained from Fermi-LAT data. The grey points represent the upper limits. The middle panel presents the X-ray light curve generated using all available observation IDs from July 2007 to January 2023. Finally, the bottom panel shows the light curves for all Swift-UVOT bands (V, B, U, UVM2, UVW1, UVW2), which were generated from the same observations as the Swift-XRT data.}
\label{fig:4C_lc}
\end{figure*}

\subsection{\it Fermi-LAT}
The {\it Fermi} $\gamma$-ray Space Telescope was launched in June 2008 aboard the Delta-II rocket. It has two instruments onboard: the Large Area Telescope (LAT) and the $\gamma$-ray Burst Monitor (GBM). LAT has a maximum effective area of 9500 $cm^2$ at normal incidence in the 1-10 GeV energy range. It has a field of view (FoV) of 2.4 sr and angular resolution at 100 MeV is < $3.5^\circ$ and < $0.15^\circ$ at 10 GeV.

Fermi-LAT started collecting data from MJD 54682.655 (4th Aug 2008). Therefore, we considered the data starting from MJD 54683 to MJD 59970 (26 January 2023). %The three sources considered for this analysis are 4C +01.02, PKS 0903-57, and S4 0954+65. 
We used the \texttt{FermiPy}\footnote{\url{https://github.com/fermiPy/fermipy}} \citep{2017ICRC...35..824W} package in Python for the analysis which leverages the Fermi Science Tools\footnote{\url{http://fermi.gsfc.nasa.gov/ssc/data/analysis/documentation/}}. The \texttt{FermiPy} package requires us to define all the analysis and data selection parameters in a \texttt{YAML} configuration file. The energy range of 100 MeV to 300 GeV  and a circular region of interest of size 10 degrees was considered for the analysis for all the sources in the same period MJD 54683-59970. Further, the data were filtered using the constraints \textit{evclass=128} and \textit{evtype=3} for \textit{gtselect}. The zenith angle cut-off of 90 degrees was chosen. The filter '\textit{(DATA\_QUAL>0)\&\&(LAT\_CONFIG==1)}' was used for the \textit{gtmktime} tool. The Galactic interstellar emission model (IEM) and isotropic diffuse emission template used for the analysis are \textit{gll\_iem\_v07.fits}\footnote{\url{https://fermi.gsfc.nasa.gov/ssc/data/access/lat/BackgroundModels.html}} and \textit{iso\_P8R3\_SOURCE\_V3\_v1.txt}\footnote{\url{https://fermi.gsfc.nasa.gov/ssc/data/access/lat/BackgroundModels.html}} respectively. 

FermiPy uses the \texttt{GTAnalysis} python class which has various functions for analyses and computations. The $\gamma$-ray light curves are generated with 7-day binning using the \texttt{lightcurve} function of \texttt{GTAnalysis} class. 

\subsection{Swift-XRT}
In November 2004, NASA launched the Neil Gehrels Swift observatory equipped with three instruments: the Burst Alert Telescope (BAT), the X-Ray Telescope (XRT), and the Ultra-violet Optical Telescope (UVOT). The XRT instrument \citep{2005SSRv..120..165B}, designed as a grazing incidence Wolter-1 telescope, boasts an impressive effective area of about 110 $cm^2$ and is capable of detecting X-rays in the energy range of 0.3 to 10 keV.\par
In this study, we made use of the Swift observations that were available in the same time frame as the Fermi observations. Specifically, we analyzed 77 Swift-XRT observations taken between MJD 54475 to 59750 for S4 0954+65, 52 observations from MJD 57197 to 59878 for PKS 0903-57, and 25 observations from MJD 54283 to 59831 for 4C +01.02.\par
To analyze the X-ray data, we focused on the energy range of 0.3-8.0 keV and used the tools from the \texttt{HEASOFT} package \footnote{\url{https://heasarc.gsfc.nasa.gov/docs/software/heasoft/}} (v6.31). Initially, we used the \texttt{xrtpipeline} package \footnote{\url{https://www.swift.ac.uk/analysis/xrt/xrtpipeline.php}} (v0.13.7) to generate clean event files (or Level 2 products) from the raw data. These Level 2 products were used to select the source and background events from circular regions with radii of 50-arcsec and 100-arcsec, respectively, in the DS9 viewer. The next step involved extracting the light curves and spectrum from the source and background events using the \texttt{xselect} (v2.5b) tool \footnote{\url{https://heasarc.gsfc.nasa.gov/docs/software/lheasoft/ftools/xselect/index.html}}. During the analysis of the XRT data, we observed that six observations of PKS 0903-57 and 23 observations of S4 0954+65 had piled-up events. These observations were promptly corrected before generating the XRT products. To solve this problem, we used an annulus region instead of a circular region to extract the light curve and spectrum. The annulus region allowed us to exclude the core of the point spread function (PSF), which was involved in the piled-up events.\par
The redistribution matrix file (RMF) was obtained from the HEASARC calibration database\footnote{\url{https://heasarc.gsfc.nasa.gov/docs/heasarc/caldb/caldb_supported_missions.html}}. The ancillary response files (ARF) were generated using the \texttt{xrtmkarf} tool\footnote{\url{https://www.swift.ac.uk/analysis/xrt/arfs.php}}. While generating the ARF file, the PSF was set to yes, which allowed the creation of an ARF-corrected file to compensate for the loss of counts resulting from the annulus exclusion. The tool \texttt{grppha}\footnote{\url{https://heasarc.gsfc.nasa.gov/ftools/caldb/help/grppha.txt}} was used to combine the source and background event files, the auxiliary response file, and the redistribution matrix file into a single spectrum file. The spectrum was fitted with a simple power-law model $ F(E) = K E^{\Gamma_x}$ using \texttt{XSPEC} (v12.13.0c)\footnote{\url{https://heasarc.gsfc.nasa.gov/xanadu/xspec/}}. During the fitting process, we used $N_H$\footnote{\url{https://www.swift.ac.uk/analysis/nhtot/}} values \citep{ahnen2018detection, shah2021unveiling, malik2022multiwavelength} of 5.17$\times 10^{20} cm^{-2}$, 2.6$\times 10^{21} cm^{-2}$, and 2.25$\times 10^{20} cm^{-2}$ for S4 0954+65, PKS 0903-57, and 4C +01.02, respectively. %The results of the power-law fit can be seen in Tables A1, A2, and A3 for sources PKS 0903-57, 4C +01.02, and S4 0954+65 respectively.
\subsection{Swift-UVOT}
The Swift Observatory's Ultra-violet Optical Telescope (\citealt{roming2005swift}) has the capability of detecting UV and optical radiation simultaneously. We acquired data in three optical filters (V, B, and U bands) and three UV filters (W1, M2, and W2 bands) over the same period in which the Swift-XRT observations were obtained. The \texttt{uvotsource} tool was used to obtain the magnitudes. Then the magnitudes were corrected for galactic extinction $A_\lambda $ and reddening by considering a value of E(B-V) = 0.397 (\citealt{schlafly2011measuring}) and calculated the $A_\lambda$ (\citealt{giommi2006swift}) for all filters used. To obtain flux measurements, we converted the corrected magnitudes using appropriate conversion factors, dust absorption relation $Mag_0 = Mag - A_\lambda$, and zero-point magnitude relation (\citealt{breeveld2011updated}) given as $Z_{pt} = -2.5log(F) - Mag_0 $, where $Z_{pt}$ is Zero point magnitude and F is the flux density in the units of $ erg cm^{-2} s^{-1}$/\AA. To generate the multi-waveband light curve, we used the given procedure as above using the \texttt{uvotsource} tool. For constructing the SED, all the image files were combined using the \texttt{uvotimsum} tool.

%\begin{figure*}
%\centering
%\includegraphics[width=.33\linewidth]{Fvar_vs_logfrequency_S40954+65_modified.pdf}\hfill
%\includegraphics[width=.33\linewidth]{Fvar_vs_logfrequency_PKS0903-57_modified.pdf}\hfill
%\includegraphics[width=.33\linewidth]{Fvar_vs_logfrequency_4C+01.02_modified.pdf}

%\caption{The fractional variability was computed for each observed waveband (Table-\ref{tab: Fvar vs energy}). The left, middle, and right panel shows the variability levels for source S4 0954+65, PKS 0903-57, and 4C +01.02, respectively. } %\textbf{\textcolor{red}{a modified plot}} }
%\label{fig:Fvar vs energy}
%\end{figure*}

\begin{figure*}
\centering
\includegraphics[width=.33\linewidth]{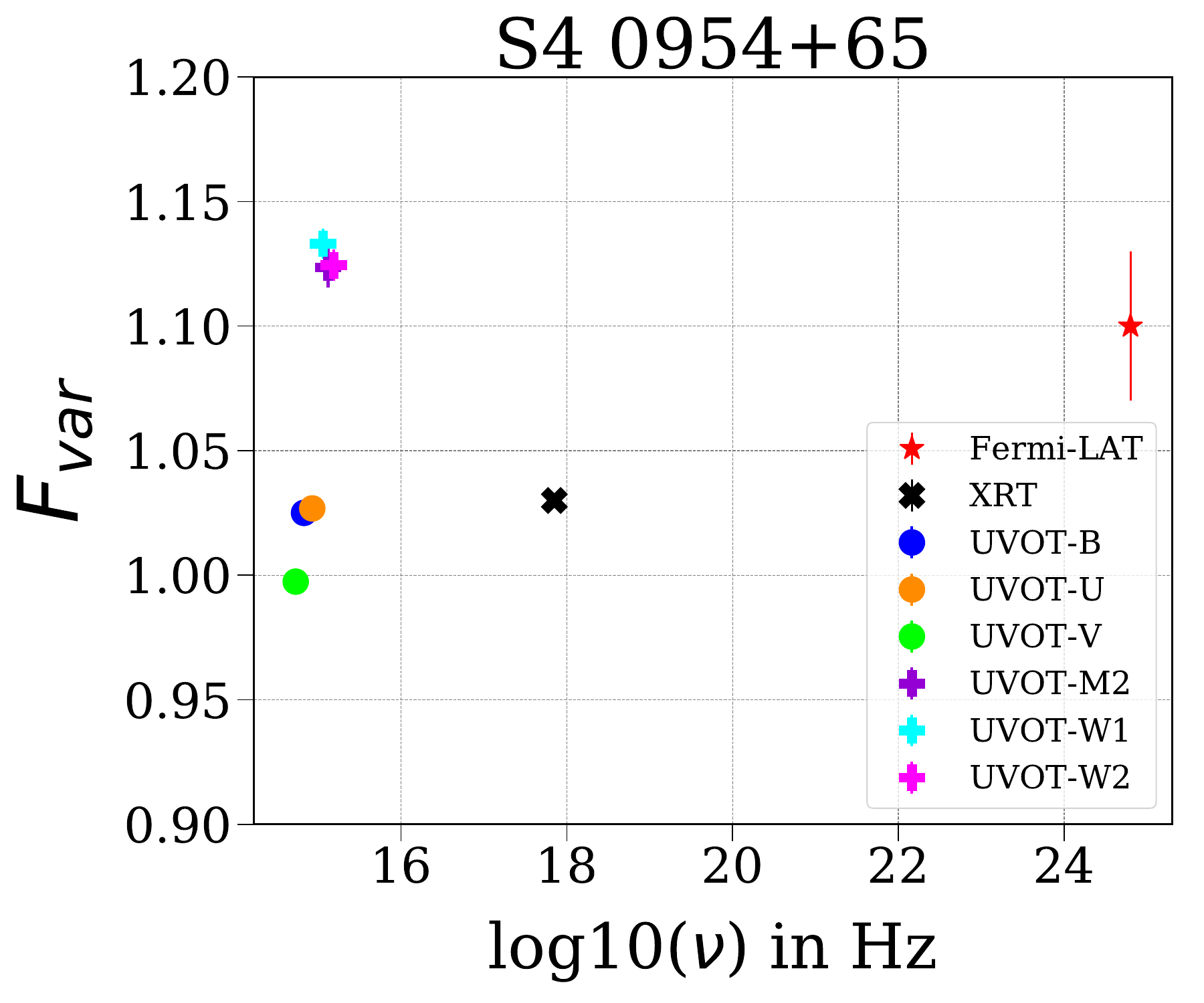}\hfill
\includegraphics[width=.33\linewidth]{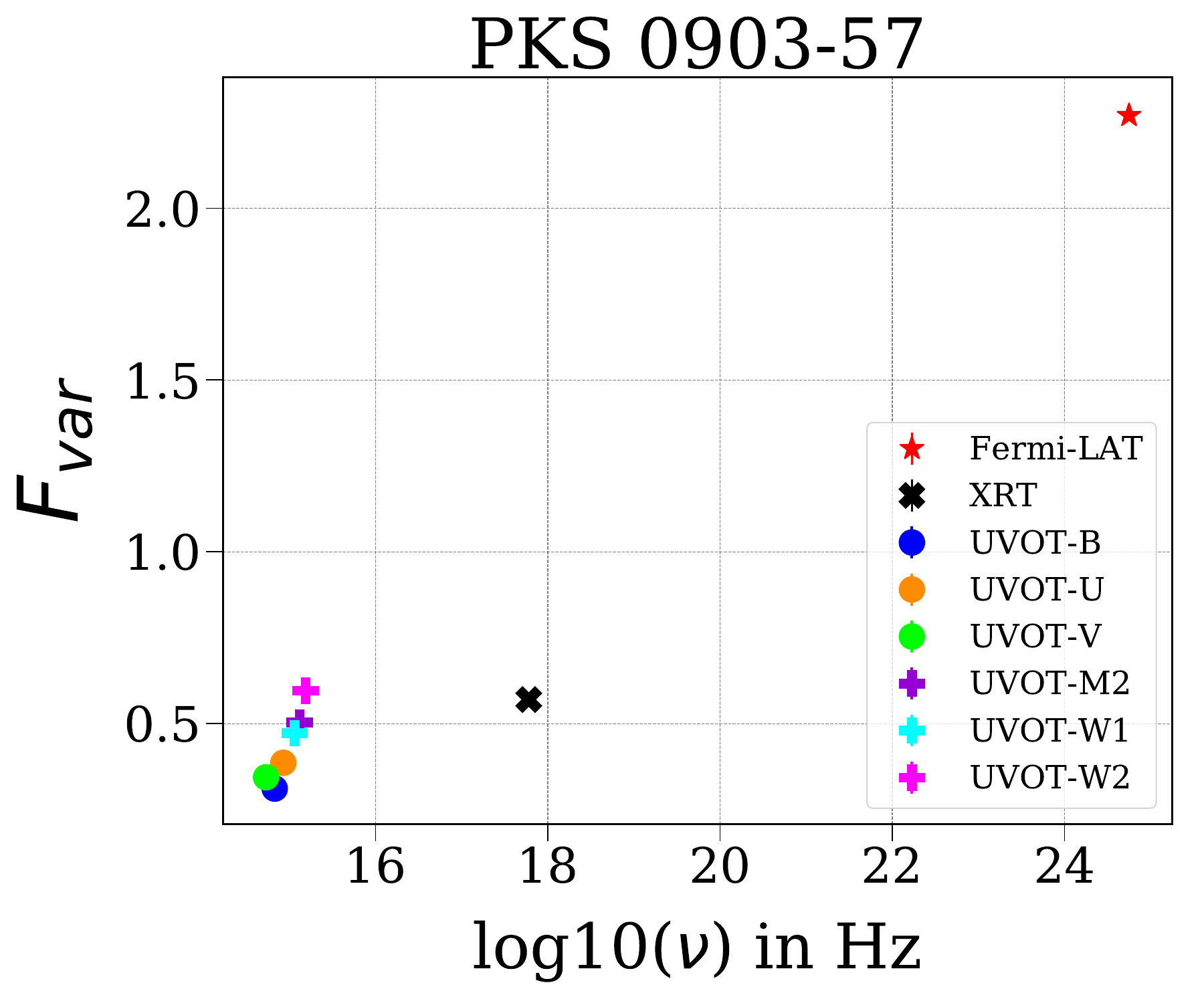}\hfill
\includegraphics[width=.33\linewidth]{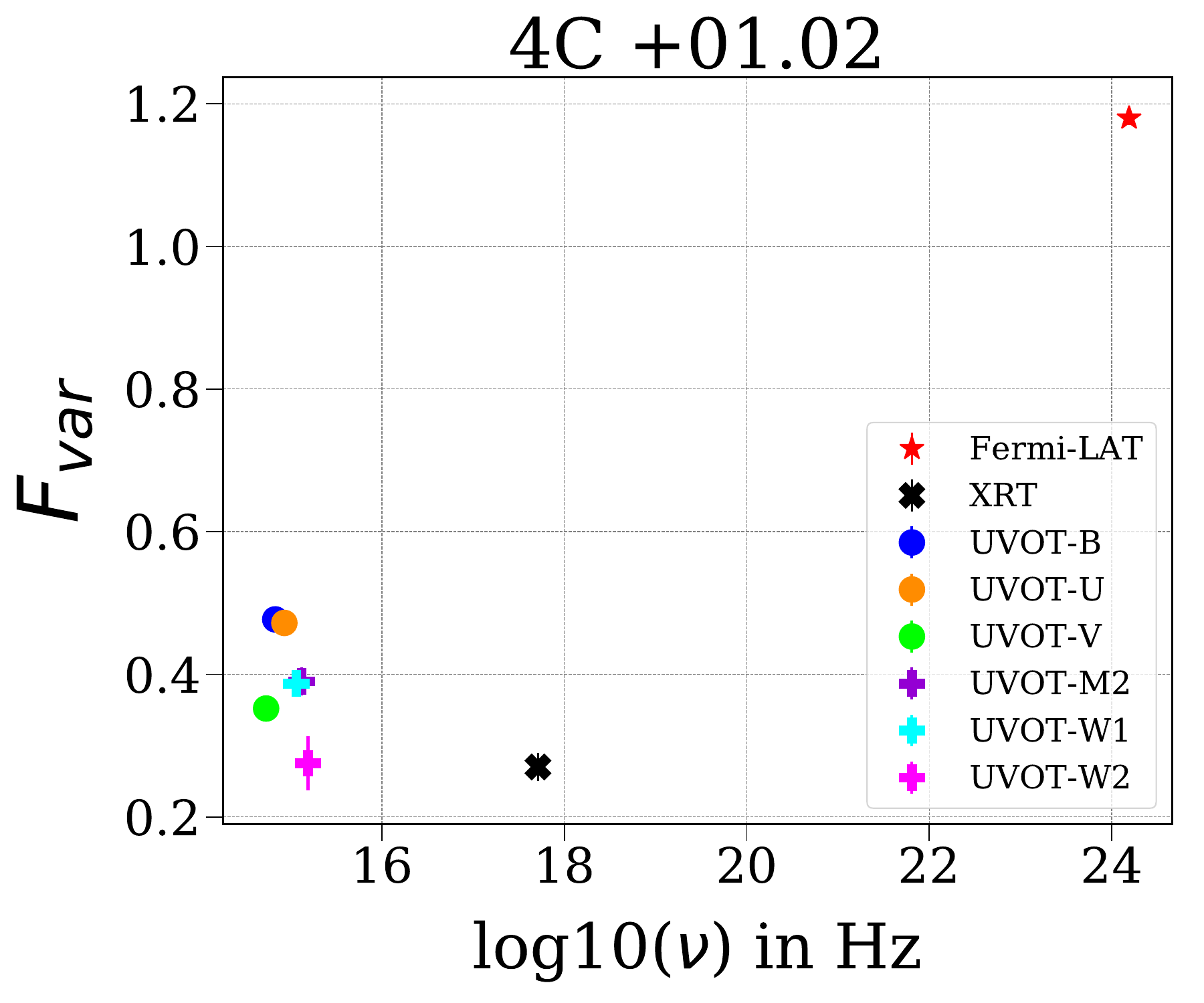}

\caption{The fractional variability was computed for each observed waveband (Table-\ref{tab: Fvar vs energy}). The left, middle, and right panel shows the variability levels for source S4 0954+65, PKS 0903-57, and 4C +01.02, respectively.} 
\label{fig:Fvar vs energy}
\end{figure*}
%{Fvar_vs_logfrequency_S40954+65_Fermiergs+without_Upperlimit_withold_xrtOLD+uvot_FINAL.pdf}
%%%%%%%%%%%%%%%%%%%%%%%   Emmanoulopoulos PSD
%\newpag
%PSD_S40954+65_significance_modified.pdf
%PSD_PKS0903-57_fb0.004_significance_modified.pdf
%PSD_4C_significance_modified.pdf

\begin{figure*}
\centering
\includegraphics[width=.32\textwidth]{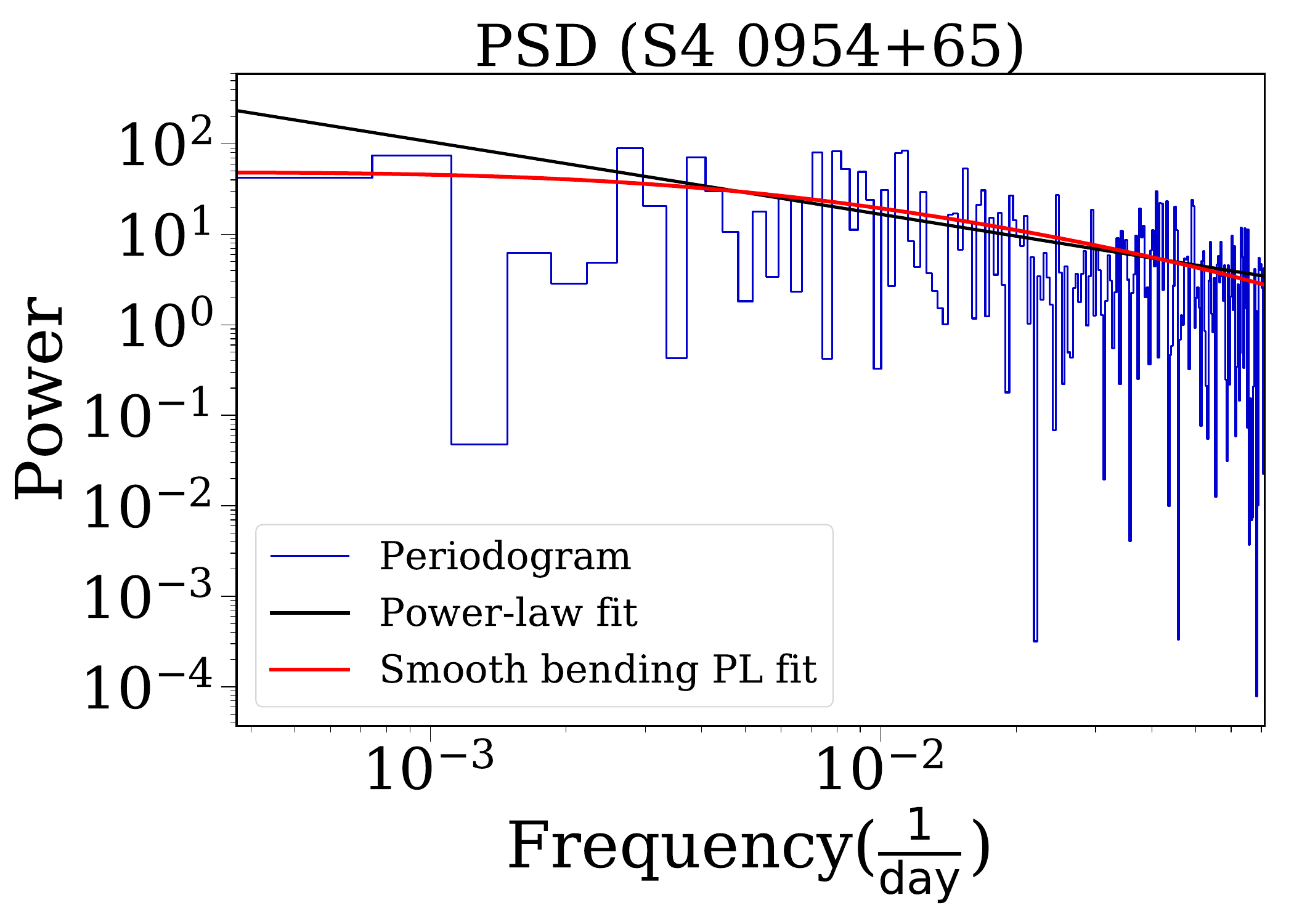}\hfill
\includegraphics[width=.32\textwidth]{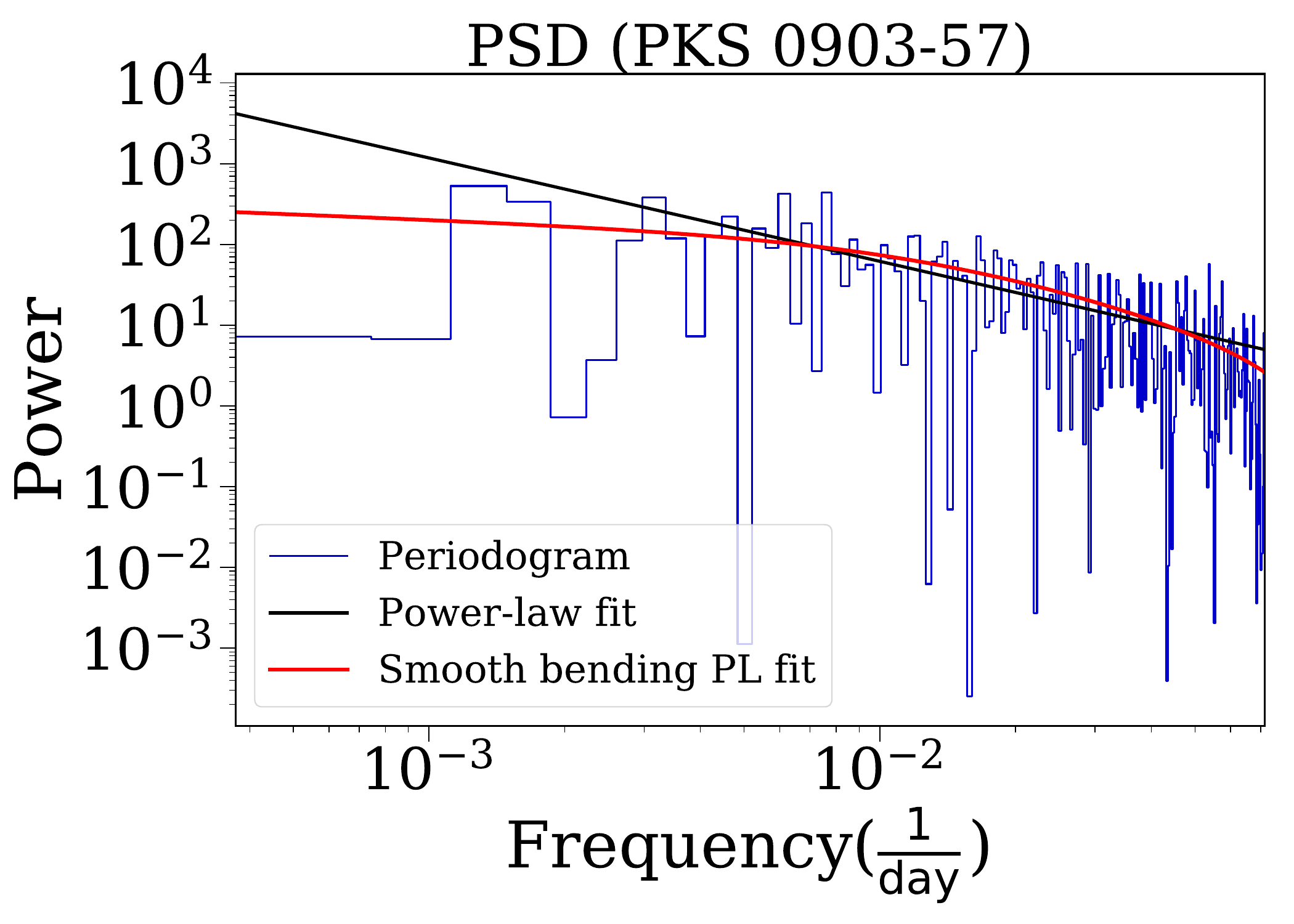}\hfill
\includegraphics[width=.32\textwidth]{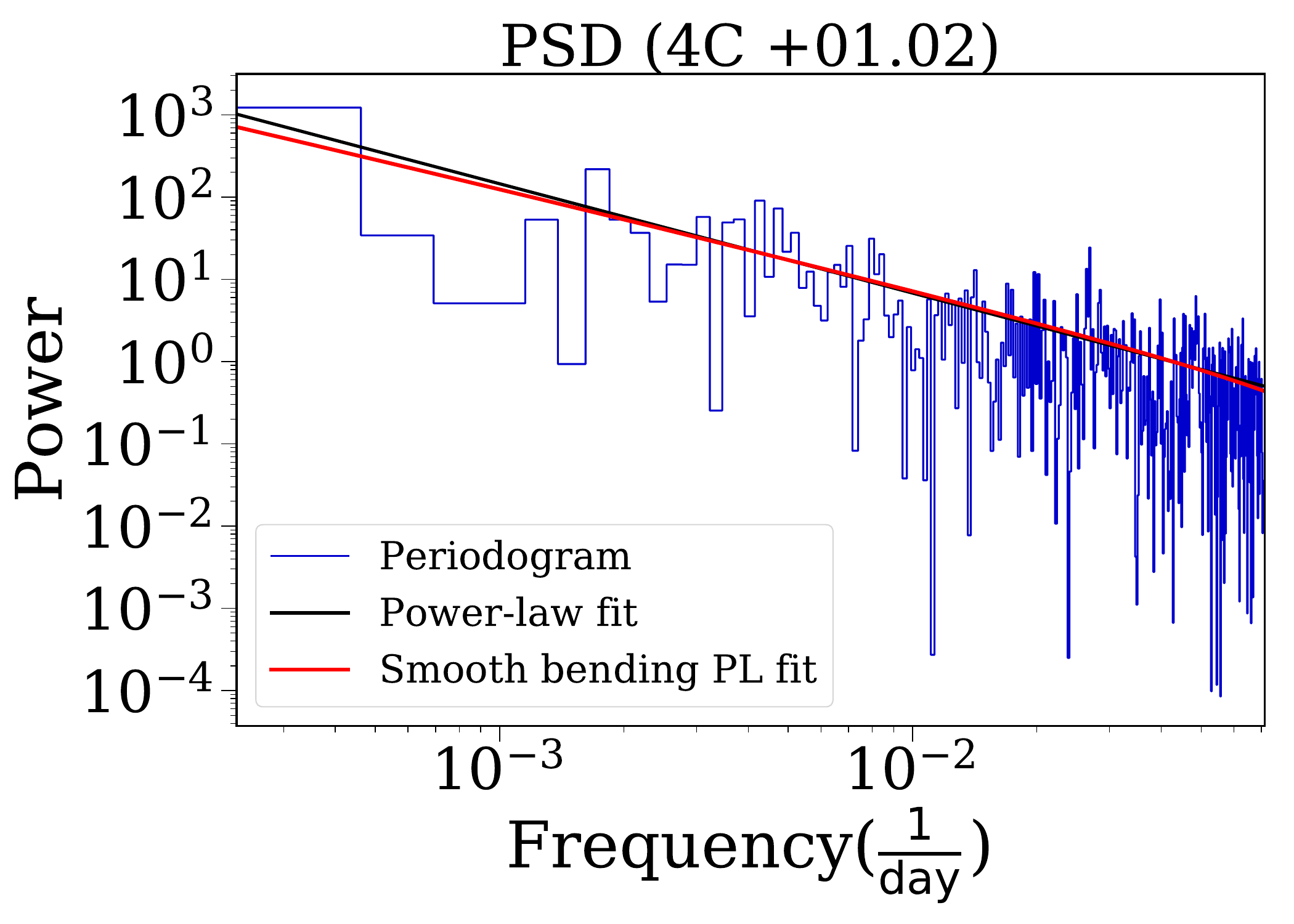}
\caption{The PSD of the $\gamma$-ray light curves, which were fitted with a power-law (black solid line) and a smooth bending power-law model (red solid line).} %Two of the PSDs (S4 0954+65 $\&$ PKS 0903-57) exhibit a break feature, which is indicated by a vertical pink line. The significance threshold at 2$\sigma$ and 3$\sigma$ levels are represented by orange and green lines, respectively.  } %\textbf{\textcolor{red}{a modified plot}}}
\label{fig:gammaPSD}
\end{figure*}

%\newpage
%%%%%%%%%%%%%%%%%%%%%%%%%%%%%%%%%%%%%%%%%%%%%%%%%%%%%%%%%%%%%
%%%%%%%%%%%%%%%%%%%%%%   CARMA modeling  %%%%%%%%%%%%%%%%%%%%
%%%%%%%%%%%%%%%%%%%%%%%%%%%%%%%%%%%%%%%%%%%%%%%%%%%%%%%%%%%%%
%Combined_DRW_fitted_LCs_residulas_ACFs_TS9_final.pdf
\begin{figure*}
\centering
\includegraphics[width=0.95\linewidth]{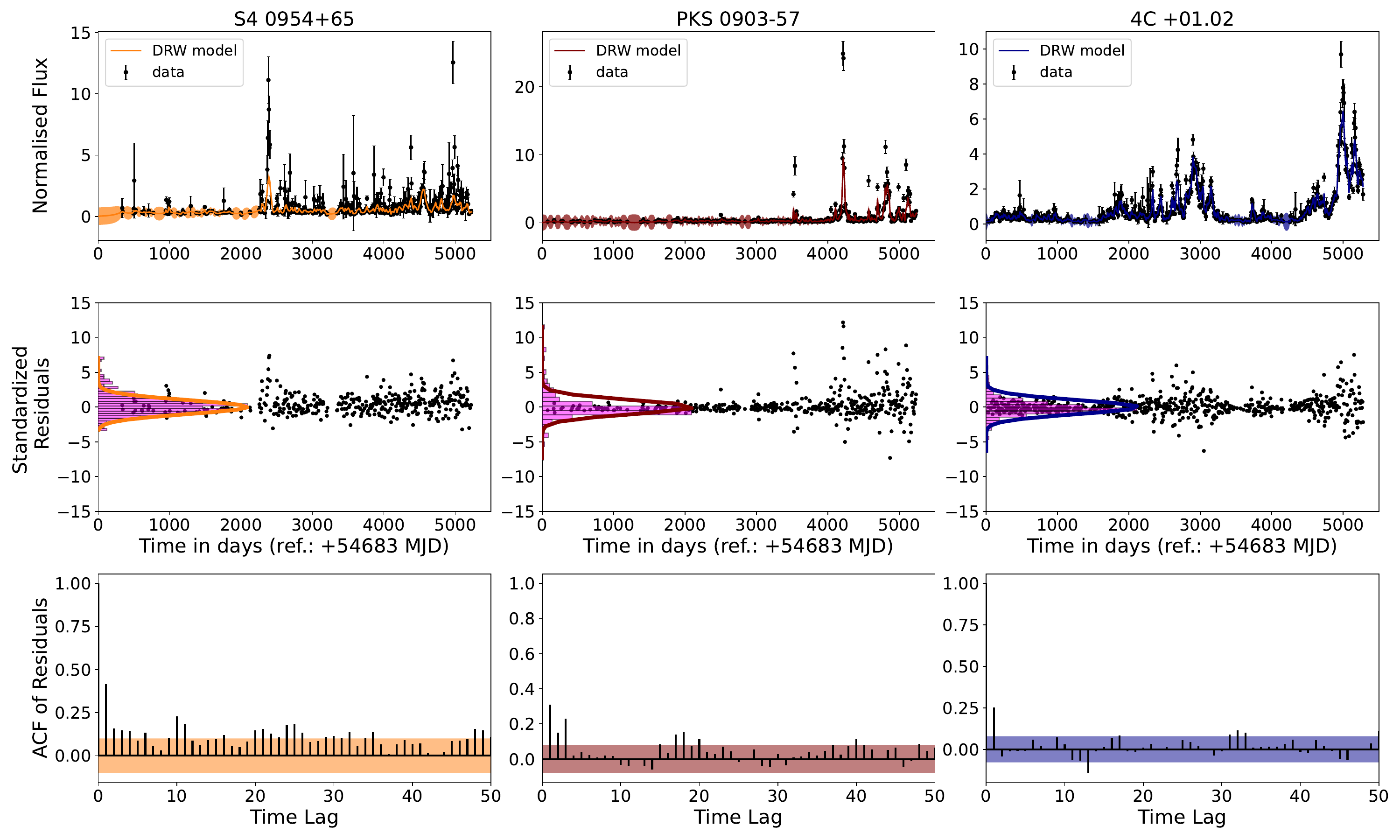}
\caption{The figure showcases the outcomes of DRW modeling applied to the 7-day binned $\gamma$-ray lightcurves of three distinct sources: S4 0954+65 (left column), PKS 0903-57 (middle column), and 4C +01.02 (right column). In the top row, the observed LAT data points (black points), with associated errors. The DRW best fits are displayed as bright orange, maroon, and blue lines, for the three sources, respectively. In the middle row, we visualize the standardized residuals of each bin, the magenta histogram shows the scaled distribution of standardized residuals, while the orange, maroon, and blue solid lines represent the scaled normal distributions in these sources. Finally, the bottom row features the autocorrelation functions (ACFs) of the residuals with 95$\%$ confidence limits of the white noise.}
\label{fig:DRW_fitted_LC}
\end{figure*}

\begin{figure*}
\centering
\includegraphics[width=0.7\linewidth]{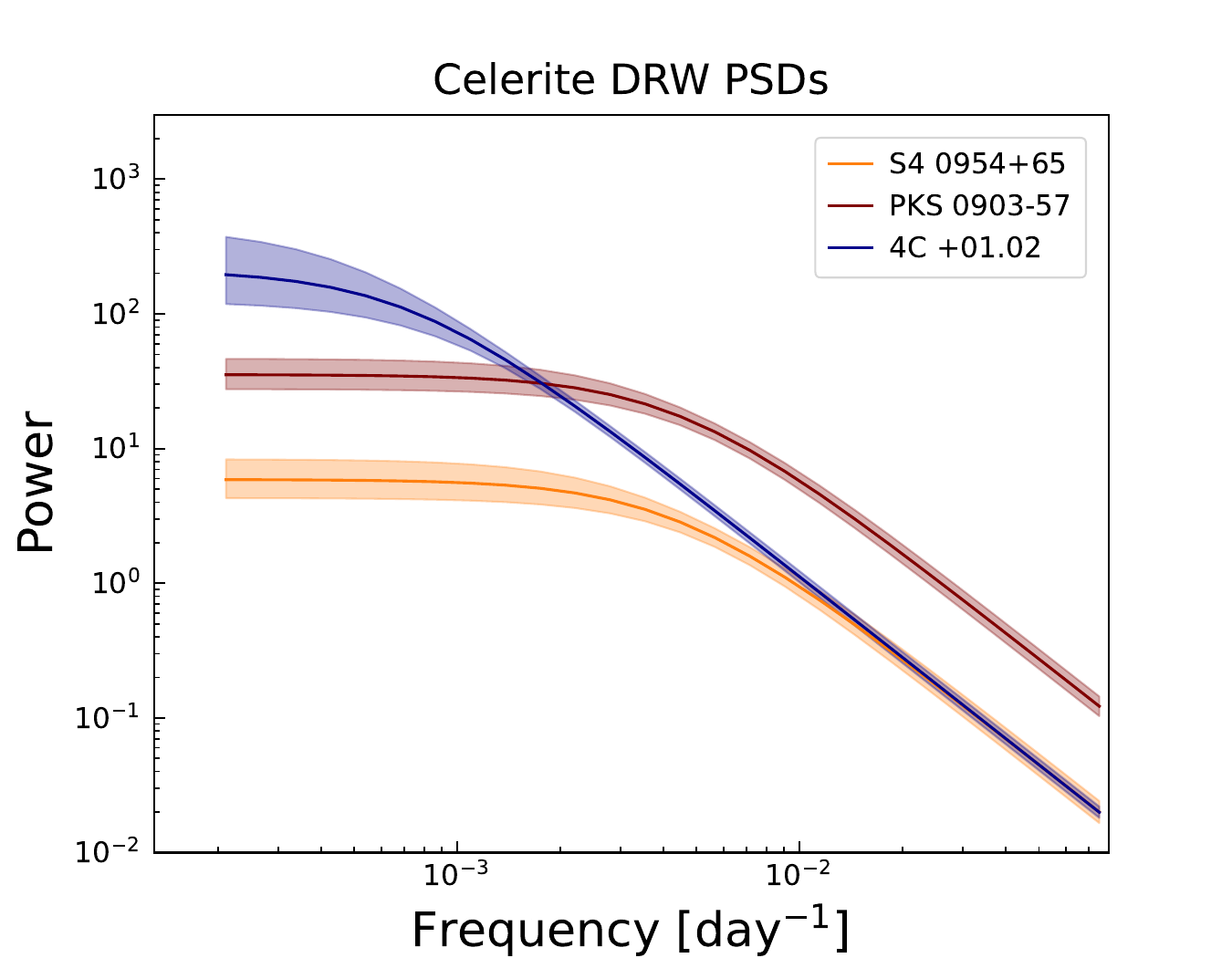}
\caption{The PSDs of $\gamma$-ray light curves constructed from the Celerite DRW modeling results for three sources are depicted in this figure: S4 0954+65 (in orange), PKS 0903-57 (in brown), and 4C +01.02 (in blue). The shaded regions in corresponding colors encompass the 1$\sigma$ confidence intervals.}
\label{fig: combined_DRW_PSD}
\end{figure*}

%%%%%%%%%%%%%%%  DRW parameters %%%%%%%%%%%%%%%%%%%%
\begin{figure*}
\centering
\includegraphics[width=.33\linewidth]{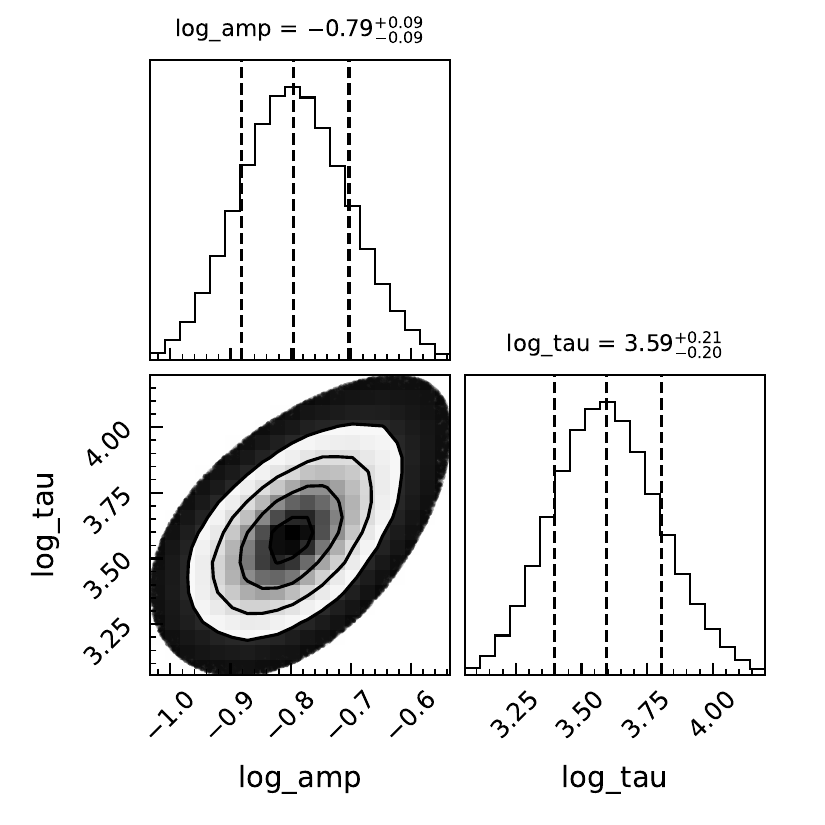}%\hfill
\includegraphics[width=.33\linewidth]{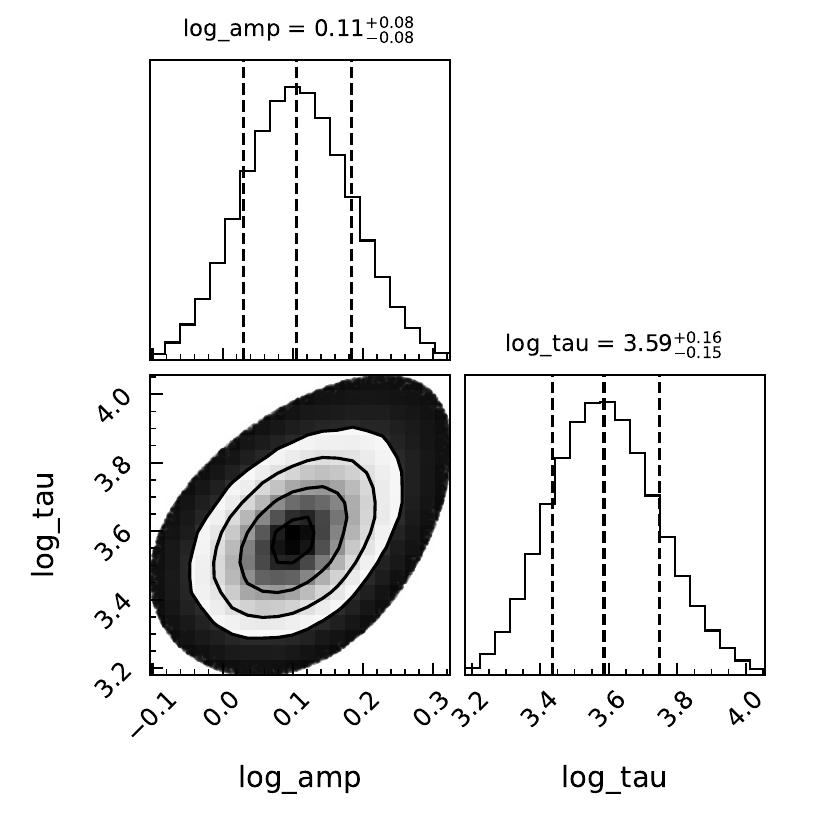}%\vfill
\includegraphics[width=.33\linewidth]{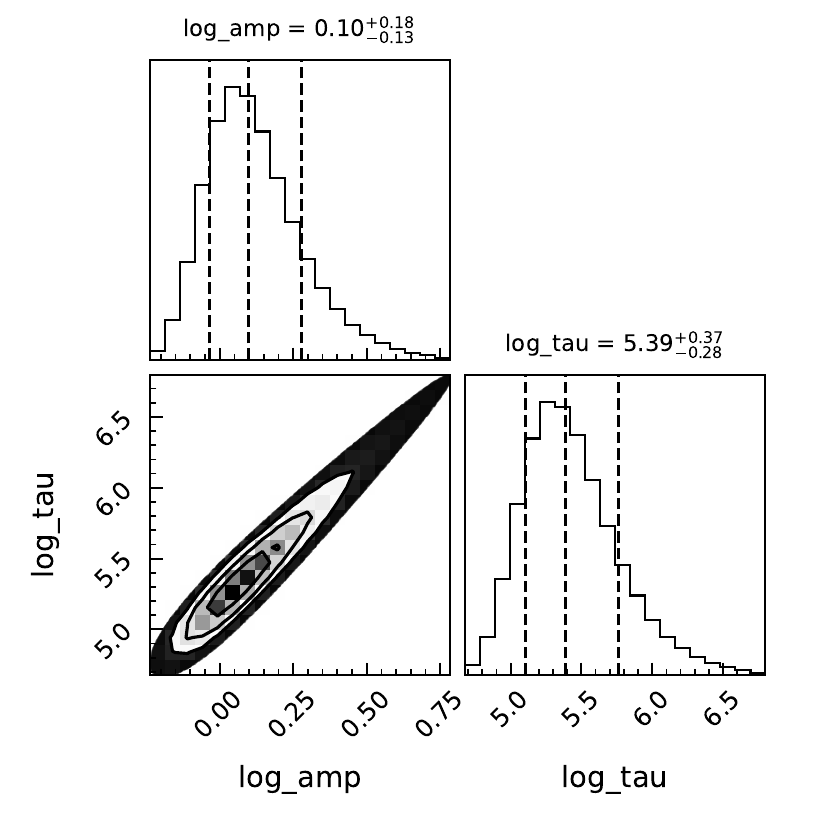}
\caption{The posterior probability densities of the DRW model parameters for all three blazars. The left panel corresponds to source S4 0954+65, the middle panel represents PKS 0903-57, and the right panel pertains to 4C +01.02.}
\label{fig: Parameters from DRW model}
\end{figure*}

%%%%%%%%%%%%%%%%%%%%%%%%%%%%%%%%%%%%%%%%%%%%%%%%%%%%%%%%%%%%%%%%
%%%%%%%%%%%%%%   Fvar v/s Energy   %%%%%%%%%%%%%%%%%%%%%%%%%%%%%

%\newpage
\section{Methodologies and Results}
In order to comprehensively study the statistical variability characteristics of the selected blazars, an in-depth study of the $\gamma$-ray light curve was carried out by applying various analysis methods including fractional variability, PSD analysis, DRW modeling and flux-distribution along with extensive Monte Carlo (MC) simulations. The details of the analysis methods and corresponding results are presented in the following sections.

%\subsection{MULTI-WAVELENGTH VARIABILITY}
%\subsection{\textcolor{red}{Variability Study}}
%\textcolor{red}{
%Our investigation employs various methodologies for conducting a comprehensive variability study. These methodologies encompass Power Spectral Density (PSD) analysis, wherein PSDs are subjected to fitting using two distinct models – the Power-Law and the Bending Power Law. These fittings serve to unravel the underlying nature of the PSDs. Additionally, within the fractional variability analysis, we computed the variability amplitude for different energy bands. In addition, we extended our analysis by employing the damped random walk model to facilitate stochastic modeling using $\gamma$-ray lightcurves. Detailed discussion is given below:}

\subsection{\bf Fractional Variability}
%\subsubsection{Measurement of excess variance and $F_{var}$}
The long-term broadband light curves for all three sources are produced and shown in Figure~\ref{fig:S4_lc}, \ref{fig:pks09_lc}, \& \ref{fig:4C_lc}. Over the long period, multiple flaring states are observed which are also accompanied by X-ray and optical-UV.
\par
The variability intensity of a light curve $x_i$ with N data points can be quantified using its variance, which can be defined as 
\begin{equation}
    S^2 = \frac{1}{(N - 1)} \sum_{i=1}^N (\langle x \rangle - x_i)^2
\end{equation}
However, when dealing with blazar light curves, finite uncertainties $\sigma_{err, i}$ arise from measurement errors, such as Poisson noise in photon counting. These uncertainties in the individual flux measurements introduce additional variance in estimating the source's variability. To accurately assess the intrinsic variability of the source, it is crucial to account for the effect of measurement noise. This is achieved by calculating the excess variance ($\sigma_{XS}$), which represents the variance of the light curve after subtracting the contribution caused by measurement uncertainties \citep{nandra1997asca, edelson2002x}
\begin{equation}
    \sigma_{XS}^2 = S^2 - \overline{\sigma_{err}^2},
\end{equation}
where, $\overline{\sigma_{err}^2}$ is the mean square error, which can be defined as 
\begin{equation}
    \overline{\sigma_{err}^2} = \frac{1}{N} \sum_{i=1}^N \sigma_{err,i}^2
\end{equation}
The excess variance serves as an estimator of the intrinsic source variance. Additionally, the fractional variability amplitude (\textit{$F_{var}$}) is defined as the square root of the normalized excess variance, which is given as, $\sigma_{NXS}^2 = \frac{\sigma_{XS}^2}{\overline{x}^2}$. It measures the level of intrinsic variability relative to the square of the mean flux measurement, providing valuable insights into the source's variability characteristics. The expression for \textit{$F_{var}$} \citep{vaughan2003characterizing} is given as 
\begin{equation}
    F_{var} = \sqrt{\frac{S^2 - \overline{\sigma_{err}^2}}{\overline{x}^2}}
\end{equation}
The parameter $F_{var}$ provides a measure of the percentage of variability amplitude and is determined by a linear statistic. %Estimating the error on $F_{var}$ using the standard error propagation formula assumes uncorrelated Gaussian processes, which may not hold true for blazar light curves. These light curves exhibit strong correlations and non-Gaussian features \citep{vaughan2003characterizing}. To address this, a Monte Carlo approach was employed to account for the impact of measurement errors on the estimation of $F_{var}$ and $\sigma_{NXS}^2$. This involved generating red noise light curves and incorporating Poisson noise. The excess variance was then recorded through multiple iterations of this process. By analyzing the distribution of these excess variance values, the uncertainty in the variance resulting from the Poisson noise could be estimated. 
The uncertainty on excess variance (\citealt{vaughan2003characterizing}) is given by
\begin{equation}
    err(\sigma_{NXS}^2) = \sqrt{\left(\sqrt{\frac{2}{N}} \cdot \frac{\overline{\sigma_{err}^2}}{\overline{x}^2}\right)^2 + \left(\sqrt{\frac{\overline{\sigma_{err}^2}}{N}} \cdot \frac{2F_{var}}{\overline{x}} \right)^2 } 
\end{equation}
and the uncertainty of \textit{$F_{var}$} (\citealt{poutanen2008superorbital, bhatta2018microvariability}) is given by 
\begin{equation}
    \Delta F_{var} = \sqrt{F_{var}^2 + err\left( \sigma_{NXS}^2 \right)} - F_{var}
\end{equation}
%The equations presented above allow for the characterization of variability amplitude in light curves with uncertainties arising from both Gaussian and Poisson distributions. By considering ($F_{var}$ + $\Delta F_{var}$), the analysis accounts for fluctuations or uncertainties in flux measurements. It should be noted that a negative intrinsic excess variance, or an imaginary value of $F_{var}$, suggests no variability in the light curve or potential overestimation of measurement errors. These instances should be excluded from the analysis.\par
The fractional variability amplitude was computed across all observed wavebands, and obtained Fvar values are included in Table \ref{tab: Fvar vs energy}. The Fvar is plotted as a function of energy in Figure \ref{fig:Fvar vs energy}.
%Additionally, the measurement of $F_{var}$ is influenced by the temporal binning or sampling rate of the light curve. For light curves with smaller temporal bins (i.e., denser sampling), the estimated value of $F_{var}$ can be higher compared to the calculated value for light curves with larger temporal bins. This discrepancy arises because dense sampling tends to smooth out short-term variability, resulting in a potentially higher estimated $F_{var}$ value.
 %The optical-UV appears to be more variable in S4 0954+65 followed by X-ray and $\gamma$-ray.
The UV and $\gamma$-ray emission appear to be more variable in S4 0954+65 compared to optical and X-rays. In PKS 0903-57 and 4C +01.02 the $\gamma$-ray emission is highly variable compared to optical-UV and X-ray emission.
 %In PKS 0930-57, a range of variability is seen in optical-UV and the source is also very variable in X-ray \& $\gamma$-ray. Blazar 4C+01.02 was found to be in a low variability state in optical-UV and X-ray compared to $\gamma$-ray where the variability is above 70$\%$. 
 This resembles the variability pattern in other blazars as seen in \citet{Rani2016}; \citet{Prince23}, \& \citet{Abdo_2011}. It is also suggested that in some of the cases, the variability pattern resembles the shape of the broadband SED \citep{2015A&A...576A.126A, 2015A&A...578A..22A}.

%%%%%%%%%%%%%%%%%%%%%%%
\begin{table}
\centering
\setlength{\extrarowheight}{10pt}
\begin{tabular}{cccc}
\hline
\hline
\multirow{2}{*}{\large{Wavebands}} & \multicolumn{3}{c}{\large{$F_{var}$}} \\
\cline{2-4}
& S4 0954+65 & PKS 0903-57 & 4C +01.02 \\
\hline
%{OLDdata}Fermi-LAT & 0.56 $\pm$ 0.02 & 0.53 $\pm$ 0.01 & 0.71 $\pm$ 0.006 \\
Fermi-LAT & 1.1 $\pm$ 0.03 & 2.27 $\pm$ 0.013 & 1.18 $\pm$ 0.01 \\
Swift-XRT & 1.03 $\pm$ 0.001 & 0.57 $\pm$ 0.01 & 0.27 $\pm$ 0.02 \\
UVOT-W2 & 1.12 $\pm$ 0.006 & 0.59 $\pm$ 0.008 & 0.27 $\pm$ 0.03\\
UVOT-M2 & 1.12 $\pm$ 0.008 & 0.50 $\pm$ 0.009 & 0.39 $\pm$ 0.019\\
UVOT-W1 & 1.13 $\pm$ 0.006 & 0.47 $\pm$ 0.008 & 0.38 $\pm$ 0.016\\
UVOT-U & 1.026 $\pm$ 0.004 & 0.38 $\pm$ 0.005 & 0.47 $\pm$ 0.011\\
UVOT-B & 1.025 $\pm$ 0.004 & 0.31 $\pm$ 0.005 & 0.477 $\pm$ 0.014\\
UVOT-V & 0.99 $\pm$ 0.004 & 0.34 $\pm$ 0.005 & 0.35 $\pm$ 0.015
\\[+3pt]
\hline
\end{tabular}
\caption{\label{tab: Fvar vs energy} Fractional variability ($F_{var}$) in different wavebands. The energy range for Fermi-LAT is 0.1–300 GeV and for Swift-XRT is 0.3 - 10 keV.}
\end{table}

\subsection{\bf Power Spectral Density}
%We used the notation x to represent a time series. For discrete time series, we denote the individual data points as $x_k$, measured at times $t_k$. Additionally, we used $\Delta x_k$ to represent the uncertainties associated with each measurement. For continuous time series, we use $x_t$ to represent the time series, where t$\in \mathbb{R}$.\par
The variability in the AGN light curve can also be characterized by the PSD. It determines the amplitude of variations and quantifies the power emitted at various frequencies.
%The periodogram is a widely used test for identifying periodic patterns in astrophysical time series. It provides valuable information about the power spectral density, which quantifies the power emitted at various frequencies. 
In the case of a regularly sampled time series, the PSD can be obtained by taking the modulus squared of its discrete Fourier transform (DFT). This approach allows us to analyze the frequency content of the data and determine the power associated with different frequencies. The discrete Fourier transform (DFT) of the set $\{x_k\}_{k=1,..,N}$ is
\begin{equation}
    DFT(f_s) =  \sum_{k=1}^{N} x_k exp[-2\pi i f_s t_k]
\end{equation}

where the discrete set of frequencies is given as, $f_s = \frac{s - 1}{N\delta t}, s \in {1, .. , N}$, and $\delta t = t_{k+1} - t_k $ is a constant time interval between consecutive observations. The sampling rate, denoted as SR, can be defined as the reciprocal of the time interval, denoted as $\delta t$. Mathematically, we express this relationship as $SR = 1/\delta t$. Furthermore, the Nyquist frequency, denoted as $f_{Nyq}$, is equal to half of the sampling rate, or $1/2\delta t$.
The rms-normalized PSD is defined as 
\begin{equation}
    P(f_s) = \frac{2 \delta t}{N}|DFT(f_s)|^2
\end{equation}
In the frequency range, only half of the N frequencies are physically meaningful. 
The Poisson noise floor level resulting from measurement uncertainties can be expressed as
\begin{equation}
    P_{poisson} = \frac{2 \delta t}{N}\sum_{k = 1}^N \Delta x_{k}^2
\end{equation}
%The light curves of the sources exhibit fluctuations in their flux values, indicating inherent variability. As a result, it is appropriate to investigate the Power Spectral Density (PSD) as a function of temporal frequency, as it provides insights into the distribution of variability power across different time scales.
This analysis aims to identify any notable peaks or breaks in the source PSD, which could suggest the presence of characteristic time scales specific to the system. Consequently, two distinct models \citep{timmer1995generating, emmanoulopoulos2013generating}, a smoothly broken power law and a simple power law were employed to fit the PSD and identify the characteristic features.
Both models can be expressed by the given equations,

The simple power-law model,
\begin{equation}
    P(f) = P_{norm} f^{-\beta}
\end{equation}
and, the bending power-law model,
\begin{equation}
    P(f) = \frac{P_{norm} f^{-\beta_1}}{1 + \left(\frac{f}{f_{break}} \right)^{\beta_2 - \beta_1}} + C
\end{equation}
where parameter C represents an estimation of the Poisson noise level originating from the uncertainties associated with individual measurements. The power-law model is characterized by the parameter $\beta$, while the smoothly broken power-law model involves two indices, $\beta_1$ and $\beta_2$, which correspond to low and high frequencies, respectively. The break frequency, denoted as $f_{break}$, is used to calculate the break timescale, represented as $T_{break}$, where $T_{break}$ = 1/$f_{break}$.\par
The Emmanoulopoulos algorithm\footnote{\url{https://github.com/samconnolly/DELightcurveSimulation}}, a combination of the Timmer-Koenig (TK95; \citealt{timmer1995generating}) and Schreiber \& Schmitz (SS96; \citealt{schreiber1996improved}) methods, was utilized to generate synthetic light curves by incorporating the underlying PSD and Probability Density Function (PDF) of the observed light curve. 
%This algorithm offers several advantages compared to the pure TK95 algorithm. Firstly, in our case study, a pseudo-random data set following the estimated distribution was used, rather than a shuffled version of the observed data set. Secondly, all spectral distortion effects resulting from the finite length and sampling rate of the observed data set were considered. Additionally, instead of using the original data set's phases, we substituted the Fourier phases of the TK95 products. The TK95 algorithm produces light curves that strictly adhere to a Gaussian distribution, which may not capture the diverse forms of distributions present in blazar flux. Therefore, to simulate light curves based on observed PSD and PDF, we adopted the Emmanoulopoulos method described in (\cite{emmanoulopoulos2013generating}). 
To assess the significance of the observed PSD, we employed Monte Carlo simulation. This involves generating a PSD of the synthetic light curves using the Fourier transform described above. By calculating the mean power and standard deviation of the PSD at each unique frequency, we were able to investigate the significance level of the PSD.
\par
In order to evaluate the model's goodness of fit, simulated light curves were used. These simulated light curves were rebinned to mimic the observed data and subjected to Fourier transformation in exactly the same way as the observed data. The assessment of the model's fit is based on the construction of two statistical functions \citep{Goyal_2022}, similar to $\chi^2$ statistics, defined as
\begin{equation}
    \chi_{obs}^2 = \sum_{\nu_k = \nu_{min}}^{\nu_{max}} \frac{[PSD(\nu_k)_{obs} - \overline{PSD(\nu_k)_{sim}}]^2}{(\Delta \overline{PSD(\nu_k)_{sim}})^2}
\end{equation}
and, 
\begin{equation}
    \chi_{dist,i}^2 = \sum_{\nu_k = \nu_{min}}^{\nu_{max}} \frac{[PSD(\nu_k)_{sim,i} - \overline{PSD(\nu_k)_{sim}}]^2}{(\Delta \overline{PSD(\nu_k)_{sim}})^2}
\end{equation}
In the equations, \textit{$PSD(\nu_k)_{obs}$} represents the observed PSD, while \textit{$PSD(\nu_k)_{sim,i}$} corresponds to the simulated PSD obtained from the simulated light curves. $\overline{PSD(\nu_k)_{sim}}$ denotes the mean, and $\Delta PSD(\nu_k)_{sim}$ represents the standard deviation, calculated by averaging the numerous PSDs generated. Here, the variable \textit{k} pertains to the unique frequencies within the PSD, while the index \textit{i} spans over the number of simulated light curves. The $\chi_{obs}^2$ represents the minimum $\chi^2$ value obtained when comparing the model to the actual data. The $\chi_{dist}^2$ values are used to evaluate the goodness of fit associated with the $\chi_{obs}^2$. It is important to note that neither the $\chi_{dist, i}^2$ nor the $\chi_{obs}^2$ follow a standard $\chi^2$ distribution. Hence, to determine the acceptance or rejection of a model for a specific simulated light curve, the $\chi_{dist, i}^2$ values were compared to the $\chi_{obs}^2$. To achieve this, the $\chi_{dist, i}^2$ values were sorted in ascending order. The model was considered acceptable if the $\chi_{dist, i}^2$ was greater than the $\chi_{obs}^2$. The success fraction, which indicates the goodness of fit in terms of capturing the shape and slope of the intrinsic PSD, is defined as the ratio of \textit{m} to \textit{M}. Here, \textit{m} represents the count of the number of $\chi_{dist, i}^2$ values that are greater than the $\chi_{obs}^2$ value, while \textit{M} denotes the total number of PSDs obtained from the simulated light curves. The success fraction is also commonly referred to as the \textit{$p_{\beta}$} (\citealt{uttley2002measuring, chatterjee2008correlated}). A high value of \textit{$p_{\beta}$} or a large success fraction indicates a good fit to the PSD, indicating that a substantial proportion of randomly generated realizations capture the desired characteristics of the intrinsic PSD and providing insight into the likelihood of a model being accepted.

 Following the above steps we produced the PSD of all three blazars for $\gamma$-ray, X-ray, and optical/UV. Due to the bad sampling of data in X-ray and optical/UV, we could not recover the good PSD, and the results are not shown here. For $\gamma$-ray, the PSD is generated in the frequency range of $10^{-4} - 10^{-1}$ day$^{-1}$ and fitted with both a power-law and bending power law model. The fitted parameters and the corresponding plots are shown in Table~\ref{tab: PSD fitting} and Figure~\ref{fig:gammaPSD}. We have computed the Akaike Information Criterion (AIC) and Bayesian Information Criterion (BIC) values to determine the best-fit model for the PSD of all sources. The results are detailed in Table~\ref{tab: PSD fitting}. It is evident from the AIC or BIC values that the bending power-law model provides the best fit for the PSD of all sources. In addition, We also estimate the \texttt{success fraction (SF)} in order to find out the goodness of the fit. A high value of \texttt{SF} indicates a good fit.
 %, signifying that a significant number of random realizations capture the shape and slope of the intrinsic (input) PSD.} 
 We noticed that the bending power-law model provides the most favorable fit to the PSDs of all the sources. Moreover, the outcomes of the success fraction provide additional support for this conclusion. \\
 %\sout{We noticed that blazar 4C+01.01 have a higher SF for power law than bending power law suggesting power law is the best fit for the PSD with a slope of 1.35$\pm$0.18. A similar power law PSD slope is also seen for other blazars 3C 279, PKS 1510-089, and PKS 0735+178 in the literature, a study done by} \citep{goyal2017multiwavelength, Goyal_2022}. \\
In the case of blazar S4 0954+65, %the PSD exhibited a good fit with the bending power-law model, yielding 
we observed a PSD break of 212 days  ($\sim$0.0047 $\mathrm{days}^{-1}$). Similarly, for blazar PKS 0903-57 and 4C +01.02, %the bending power-law model emerged as the appropriate choice for characterizing the break in the PSD, 
the PSD break is observed at 256 days ($\sim$0.0039 $\mathrm{days}^{-1}$) and 714 days ($\sim$0.0014 $\mathrm{days}^{-1}$).
%\sout{For blazar, S4 0954+65, the SF for both power law and bending power-law PSD is almost similar with a power law slope of 0.95. With a bending power-law model, it shows a break at $\sim$ 152 days (frequency = 0.0066 day$^{-1}$). For the blazar PKS 0903-57 the power law gives a small SF of 26$\%$ and the bending power law gives SF of 66$\%$ suggesting the bending power law as the best PSD model and showing a break at frequency 0.004 day$^-1$ or a time scale of 250 days.} 
The break in the high-energy PSD is observed only in a handful of blazars before \citep{ryan2019characteristic}. The details about the other fitted parameters are shown in Table ~\ref{tab: PSD fitting}.

%%%%%%%%%%%% PSDs best fitting parameters
%\begin{table*}
%\setlength{\extrarowheight}{3pt}
%\begin{tabular}{|c|c|c|c|c|}
%\hline
%Model & Parameter & S4 0954+65 & PKS 0903-57 & 4C +01.01 \\
%\hline
%\multirow{2}{*}{Power-law} & Normalisation constant & 0.24 & 0.09 & 0.01 \\
%& Slope & 0.95 $\pm$ 0.2 & 1.24 $\pm$ 0.27 & 1.35 $\pm$ 0.18 \\
%\cline{2-5}
%& Success-fraction & 71\% & 26.2\% & 68.6\% \\
%\cline{2-5}
%& \textbf{AIC} & \textbf{4230.47} & \textbf{5430.64} & \textbf{4546.98} \\
%\cline{2-5}
%& \textbf{BIC} & \textbf{4238.31} & \textbf{5438.49} & \textbf{4554.84} \\
%\hline
%\multirow{5}{*}{Bending power-law} & Normalisation constant & 47.5 & 15.44 & 0.007 \\
%& Low frequency index & 0.1 & 0.43 & 1.42 \\
%& High frequency index & 1.48 & 1.29 & 3.82 \\
%& Break frequency & 0.0066 & 0.004 & 0.28 \\
%& Offset & 1.22 & -2.4 & 0.12 \\
%\cline{2-5}
%& Success-fraction & 72\% & 66\% & 55\% \\
%\cline{2-5}
%& \textbf{AIC} & \textbf{3050.18} & \textbf{4023.81} & \textbf{4791.79} \\
%\cline{2-5}
%& \textbf{BIC} & \textbf{3069.79} & \textbf{4043.43} & \textbf{4811.45} \\
%\hline
%\end{tabular}
%\caption{\label{tab: PSD fitting} Best-fit parameters obtained from fitting power-law and bending power-law models to the power spectral densities (PSDs) derived from $\gamma$-ray light curves of all sources.}
%\end{table*}

\begin{table*}
\setlength{\extrarowheight}{3pt}
\begin{tabular}{|c|c|c|c|c|}
\hline
Model & Parameter & S4 0954+65 & PKS 0903-57 & 4C +01.01 \\
\hline
\multirow{2}{*}{Power-law} & Normalisation constant & 0.41 & 0.16 & 0.01 \\
& Slope & 0.80 & 1.28 & 1.32 \\
\cline{2-5}
& Success-fraction & 48.4$\%$ & 24.7$\%$ & 40$\%$ \\
\cline{2-5}
& AIC & 1691.80 & 2794.40 & 3576.93 \\
\cline{2-5}
& BIC & 1698.32 & 2800.91 & 3584.39 \\
\hline
\multirow{5}{*}{Bending power-law} & Normalisation constant & 0.28 & 265.37 & 0.14 \\
& Low frequency index & 1.02 & 0.03 & 1.07 \\
& High frequency index & -0.04 & 1.22 & 1.35 \\
& Break frequency (1/day) & 0.0047 & 0.0039 & 0.0014 \\
& Offset & -1.15 & -6.13 & -0.17 \\
\cline{2-5}
& Success-fraction & 73.8$\%$ & 54$\%$ & 53.7$\%$ \\
\cline{2-5}
& AIC & 1254.15 & 1838.52 & 3371.2 \\
\cline{2-5}
& BIC & 1270.43 & 1854.81 & 3390.58 \\
\hline
\end{tabular}
\caption{\label{tab: PSD fitting} Best-fit parameters obtained from fitting power-law and bending power-law models to the PSDs derived from $\gamma$-ray light curves of all sources.}
\end{table*}

%%% S4 - 0.11, 1.25, 0..13, 0.008, 0.046, 1236.06, 1252.34
%%% 4c - 0.05,1.22,1.23,0.004,-0.22, 3402.96, 3421.63

\subsubsection{\bf PSD Interpretations}
 The break in the PSD is generally seen in galactic X-ray binaries and in the AGN where the emission is dominated by an accretion disc. A detailed PSD analysis has been done in a few blazars but none of them showed a break in PSD as it is expected since the emission is mostly dominated by the jet emission \citep{goyal2018stochastic, Goyal_2022}. However, in X-ray, \citet{chatterjee2018possible} have observed a break in PSD of Mrk 421 derived using the long-term as well as short-term X-ray light curve. In our sample of three sources, we observed a clear break in one of the blazars PKS 0903-57 with a break time scale of 256 days. The break in PSD is generally associated with the characteristic time scale in the light curves. In the case of a blazar, this time scale can be associated with the non-thermal emission processes in the jet such as the particle cooling, light crossing or escape time scale \citep{kastendieck2011long, sobolewska2014stochastic, finke2014fourier,chen2016particle, chatterjee2018possible, ryan2019characteristic}. However, generally in the blazar scenario, these time scales are much shorter than our observed break time scale as also discussed in \citet{chatterjee2018possible}. An alternative scenario is that the PSD break time scale is associated with the accretion process or the time scale in the accretion disc which gets propagated to the jet \citep{isobe2014maxi}. In this scenario, the time scale would be the dynamical time scale or a viscous or thermal time scale in the disc. If we believe this is the possible scenario then we should expect to see a log-normal flux distribution as we generally see in AGN \citep{vaughan2003characterizing} or X-ray binary systems \citep{Uttley_2001}. We produced the $\gamma$-ray flux distribution of PKS 0903-57 and found that it can be well-fitted with a double-lognormal model rather than a single log-normal flux distribution. More details are provided in Section 3.4. Further, we estimated the time scale associated with the disc variability such as dynamical timescale, thermal timescale, and viscous time scale. The dynamical timescale can be given as $t_{dyn}$ = 2 $\times$ 10$^3$ $R^{3/2}$ $M_8$ seconds, where R=$r/r_s$ is the distance of the hot flow in units of Schwarschild's radius ($r_s$) and in case of jet or hot corona it should be much closer to the central SMBH. Considering the BH is maximally spinning then the $r$=$r_s$.  Here, $M_8$ represents the mass of SMBH in units of $10^8M_{\odot}$. We did not find an estimate for the BH mass for PKS 0903-57 but considering the distribution of BH mass of all radio-loud AGN from \citet{Ghisellini2015}, we consider the most probable value of BH mass between 3-6$\times$10$^{8}$M$_{\odot}$. Considering the above values the dynamical timescale is estimated to be between 1-4 hrs which is much shorter than the observed PSD break timescale. The thermal timescale in the accretion disc is defined as the ratio of the disc's internal energy to the heating or cooling rate. In the standard accretion disc scenario, it is defined as, $t_{th}$ = $t_{dyn}/\alpha$, where $\alpha$ is the disc viscosity parameters generally considered to be 0.1 \citep{2006ASPC..350..183W}. Considering this $\alpha$ value the estimated timescale would be an order of a day which is again shorter than the observed PSD break timescale. Another important timescale is the viscous timescale which represents the rate at which matter flows through the accretion disc. In the standard disc scenario, it is defined as $t_{visc}$ = $t_{th}$ $(r/h)^2$, where $h$ represents the height of the accretion disc. In a typical scenario, the ratio $h/r$ = 0.1 \citep{2006ASPC..350..183W} or smaller than this. We considered $h/r = 0.08$ which gives the viscous timescale around 260 days, very close to the observed PSD break timescale. The consistency in the timescale suggests that the long-term variation seen in the $\gamma$-ray light curve could be linked with the viscous time scale in the accretion disc.

 To explain this time scale, next, we modeled the $\gamma$-ray light curve with a stochastic process as done for the accretion disc variability by \citet{kelly2009variations}. A detailed discussion is provided in the next section.

%%%%%%%%%%%%%%%%%%%%%%%%%    DRW modeling %%%%%%%%%%%%%%

\subsection{DRW Modeling}

In the realm of astronomical variability analysis, Gaussian process modeling of light curves has emerged as an alternative to conventional methods like Lomb-Scargle Periodogram (LSP), Weighted Wavelet Z-Transform (WWZ), and REDFIT (a FORTRAN 90 program is used to estimate red noise spectra from unevenly sampled time series, \citealt{schulz2002redfit}). Unlike traditional frequency domain-based approaches, Gaussian process modeling focuses on understanding variability in the time domain, providing a more comprehensive and nuanced perspective. One popular Gaussian process model \citep{brockwell2001continuous} is known as the Continuous Time Autoregressive Moving Average (CARMA) model.\par
CARMA(p,q) processes \citep{kelly2014flexible} are defined as the solutions to the following stochastic differential equation:\par

\begin{equation}
\begin{split}
\frac{d^p y(t)}{dt^p} + \alpha_{p-1}\frac{d^{p-1}y(t)}{dt^{p-1}}+...+\alpha_0 y(t) =\\
\beta_q \frac{d^q \epsilon(t)}{dt^q}+\beta_{q-1}\frac{d^{q-1}\epsilon(t)}{dt^{q-1}}+...+\beta_0 \epsilon(t),
\end{split}
\end{equation}

Here, y(t) is a time series, $\epsilon(t)$ is a continuous-time white-noise process, $\alpha^{*}$ and $\beta^{*}$ are the coefficients of the AR and MA models respectively. The parameters p $\&$ q determine the order of the AR and MA models respectively. The CARMA(p,q) model has proven immensely valuable in studying the $\gamma$-ray variability exhibited by blazars \citep{goyal2018stochastic, ryan2019characteristic, tarnopolski2020comprehensive, Yang_2021, zhang2022characterizing}. Specifically, when the model's parameters are configured as p=1 and q=0, i.e., a CAR(1) model \citep{kelly2009variations, brockwell2002introduction}. The CAR(1) (simply the lowest order CARMA model) is also known as the Ornstein-Uhlenbeck process and has often been referred to as a Damped Random Walk (DRW) process in the astronomical literature.\par
The CAR(1) can be described by the stochastic differential equation of the following form:\par
\begin{equation}
    \left[ \frac{d}{dt} + \frac{1}{\tau_{DRW}} \right]y(t) = \sigma_{DRW} \epsilon(t)
\end{equation}
where $\tau_{DRW}$ is the characteristic timescale of DRW process and $\sigma_{DRW}$ is the amplitude of random perturbations.\par
In this study, we utilized the stochastic process method provided by the \textbf{\texttt{EzTao}}\footnote{\url{https://eztao.readthedocs.io/en/v0.3.0/index.html}} package, which is built on top of \textbf{\texttt{celerite}}\footnote{\url{https://celerite.readthedocs.io/en/stable/}} package (a fast gaussian processes regression library) \citep{foreman2017fast}. To model the DRW process, an essential component is the covariance function. It is expressed as:
\begin{equation}
    k(t_{nm}) = a.exp(-t_{nm}c),
\end{equation}
where $t_{nm} = | t_n -t_m|$ is the time lag, and $a = 2 \sigma_{DRW}^2$ and $c = \frac{1}{\tau_{DRW}}$. The PSD for the DRW model is written as:
\begin{equation}
    S(\omega) = \sqrt{\frac{2}{\pi}} \frac{a}{c} \frac{1}{1 + (\frac{\omega}{c})^2}
\end{equation}

In the fitting process, we employed the Markov Chain Monte Carlo (MCMC) algorithm, which is implemented in the \textbf{\texttt{emcee}}\footnote{\url{https://github.com/dfm/emcee}} package. This approach allowed us to perform a robust fitting of the light curve data. In total, 50,000 samples were generated through the MCMC algorithm. To ensure reliable parameter estimation, the first 20,000 samples were considered as burn-in. From the remaining MCMC samples, we calculated the values and uncertainties of the model parameters.\par
The DRW model was applied to fit each $\gamma$-ray light curve and the result is shown in Figure~\ref{fig:DRW_fitted_LC}. To evaluate the quality of the fits, the probability densities of the standardized residuals and the autocorrelation function of the standardized residuals were analyzed (Figure~\ref{fig:DRW_fitted_LC}). %\sout{The produced PSD from the DRW modeling is shown in Figure~\ref{fig: combined_DRW_PSD} and we noticed a clear break in $\gamma$-ray PSD of blazar S4 0954+65 and PKS 0903-57. A possible break can also be expected in 4C +01.02 with a longer duration of the light curve. The break timescale in two blazars is marked with a vertical dashed line and it is consistent with the value estimated in Section 3.2.}
We have generated PSDs, including 1$\sigma$ confidence intervals based on the outcomes of DRW modeling, Figure~\ref{fig: combined_DRW_PSD}, for all sources. The break frequency is defined as the f$_b$ = 1/(2$\pi \tau_{DRW}$) and the $\tau_{DRW}$ in all the cases is found as 0.0043, 0.0043, and 0.0007 $\mathrm{day^{-1}}$ for S4 0954+65, PKS 0903-57, and 4C +01.02, respectively.  The break frequencies derived from PSD analysis for S4 0954+65 and PKS 0903-57 are consistent with results obtained from DRW modeling. However, in the case of 4C +01.02, the break frequency derived in both PSD analysis and DRW modeling is not quite consistent and it is because the light curve length is not sufficient to derive an actual PSD as can be seen in the PSD plot.
%from PSD analysis doesn't consistent with results obtained from DRW modeling.}
%Notably, A clear break of 264 days and 211 days is evident in the constructed PSD for Blazer S4 0954+65 and PKS 0903-57, respectively. A possible break can be expected in the PSD of 4C +01.02 at approximately 706 days.}
The posterior probability density distributions of the parameters obtained from the DRW modeling using the MCMC algorithm are shown in Figure~\ref{fig: Parameters from DRW model} and the best-fit values are shown in Table~\ref{tab: Posterior parameters}.

%%%%%%%%%%%   Characterstic timescale of relativistic jets 

\subsubsection{\bf{Variability or Characteristic Timescales}}
In this investigation, we have demonstrated that the DRW model can effectively capture $\gamma$-ray variability. By employing the DRW model for fitting the light curve, we have extracted the characteristic timescales also known as the damping time scale for the three studied sources along with their corresponding errors within a 1$\sigma$ range. The values are detailed in Table \ref{tab: Posterior parameters}.\par
However, it's worth noting that the measurement of damping timescales can be influenced by the limited length of the light curve (LC). Prior studies \citep{kozlowski2017limitations, suberlak2021improving} have indicated potential biases in these measurements due to this limitation. Moreover, \citet{Burke_2021} have established that reliable measurements of damping timescales require them to be significantly larger than the typical cadence of the LC.\par
In our analysis, we have adopted these criteria to ensure the reliability of damping timescale measurements. Specifically, the length of the LC should be at least 10 times that of the timescale under consideration, and the resultant timescale should surpass the average cadence of the LC. It is noteworthy that all the calculated characteristic $\gamma$-ray timescales for the examined sources adhere to these criteria, affirming their reliability. The fitting of LCs was conducted within the observed frame, and the corresponding damping timescale values presented in Table~\ref{tab: Posterior parameters} are also defined in the observed frame. However, to ascertain the timescale in the rest frame ($\tau^{\mathrm{rest}}_{\rm damping}$), it is necessary to account for the cosmological time dilation and the influence of the Doppler beaming effect \citep{zhang2022characterizing},
\begin{equation}
    \tau_{\mathrm{damping}}^{\mathrm{rest}} = \frac{\tau_{\mathrm{DRW}} \delta_{D}}{1 + z}
\end{equation}
where $\delta_{D}$ is the Doppler factor and  $z$ is the redshift of the source. Estimating the Doppler factor for AGN poses a challenging task. Various approaches are employed to derive this factor, including methods such as modeling the broadband SED, from rapid $\gamma$-ray variability timescale, X-ray, and $\gamma$-ray emissions (as demonstrated by \citealt{chen2018jet, pei2020estimation, zhang2020doppler}). Additionally, in recent work, \citet{liodakis2017f} proposed that the variability of Doppler factors can be computed by constraining the equipartition brightness temperature of radio flares. Despite these efforts, the determination of the Doppler factor remains subject to significant uncertainties.\par

\begin{figure*}
\centering
\includegraphics[width=0.45\textwidth]{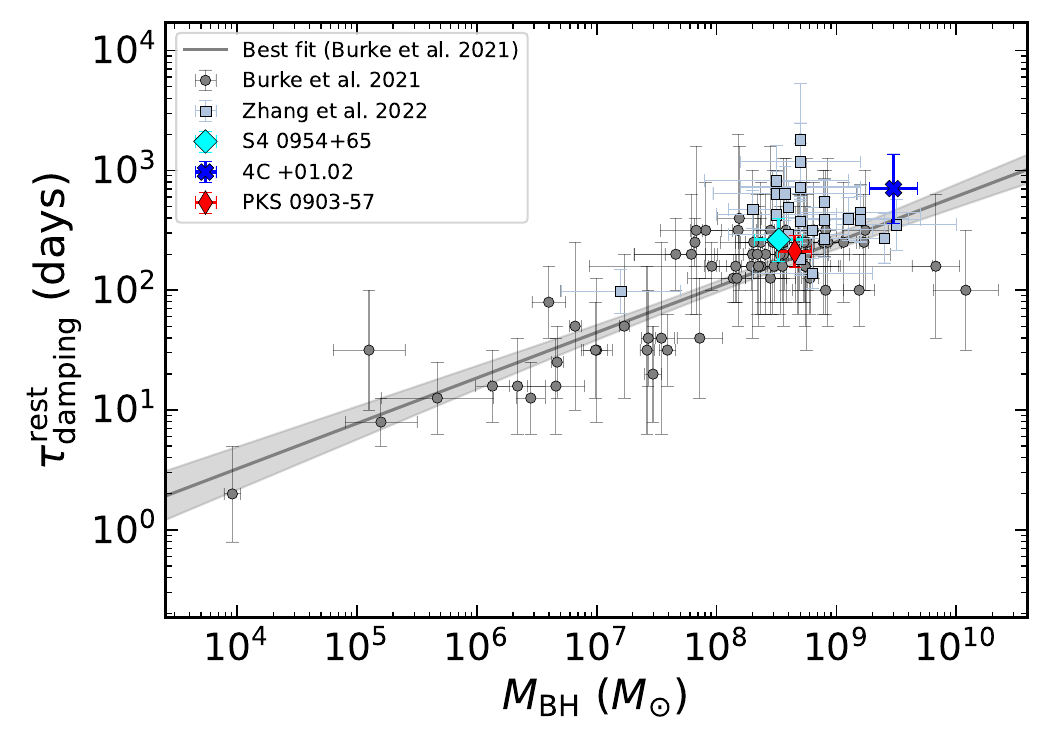}
\includegraphics[width=0.45\textwidth]{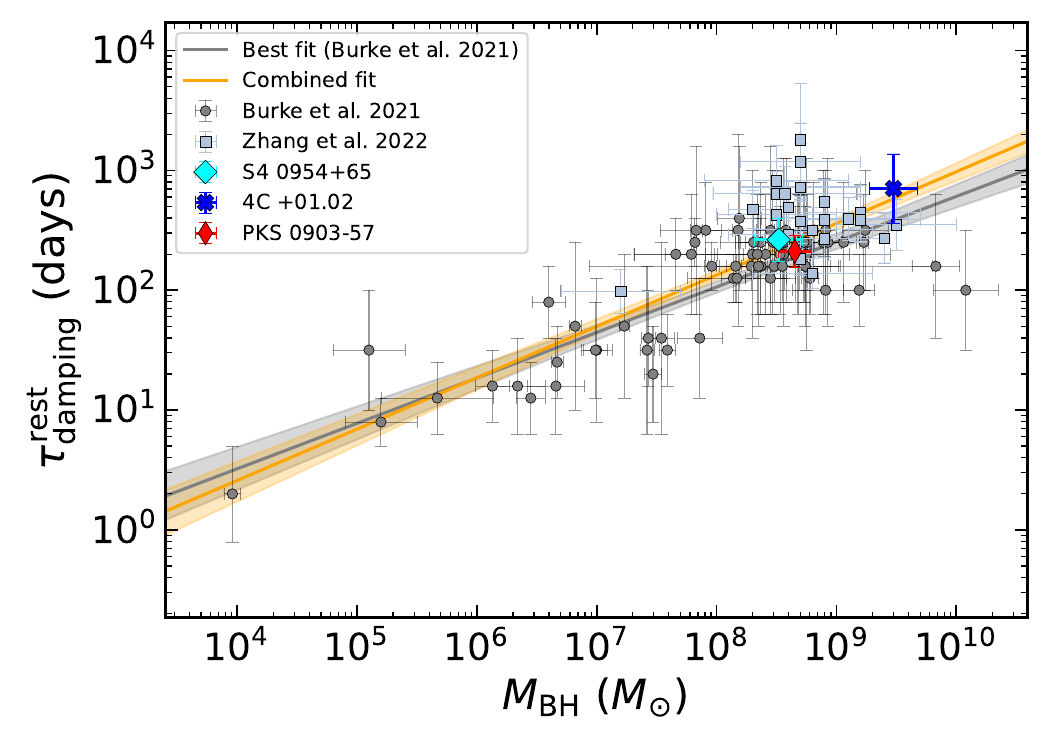}
\caption{The damping timescale in the rest frame as a function of black hole mass. In both panels, the data points, line, and area in grey color represent the optical results from \citet{Burke_2021}. The light steel blue colored data points represent the $\gamma$-ray results from \citet{zhang2022characterizing}. The cyan, blue, and red data points are the results of our $\gamma$-ray analysis, corresponding to S4 0954+65, 4C +01.02, and PKS 0903-57, respectively. In the right panel, an orange line represents a combined fit to results from our study sample, \citet{Burke_2021}, and \citet{zhang2022characterizing}.}
\label{fig:timescale vs M}
\end{figure*}

On examining these different methodologies, it becomes clear that the resulting Doppler factor values vary widely. In an attempt to reduce the uncertainty in $\delta_D$, an average value, around 10, is employed for blazars. Further investigations reveal that the rest-frame timescale ($\tau^{\mathrm{rest}}_{\rm DRW}$) of our sources spans from 150 days to 1100 days. We plot the damping timescale estimated from $\gamma$-ray for three sources in this work with optical damping timescale from \citet{Burke_2021} and $\gamma$-ray damping timescale from \citet{zhang2022characterizing} in Figure~\ref{fig:timescale vs M}. Our three sources exactly lie on the best-fit relation (grey line) from \citet{Burke_2021} suggesting that both the optical and $\gamma$-ray timescale are connected and infer to a possible connection of disc with jet.
On the right panel of Figure~\ref{fig:timescale vs M}, we fit our observed $\gamma$-ray rest-frame timescales with \citep{Burke_2021, zhang2022characterizing}, and the best fit is shown in the yellow band. The best-fit relation for the combined fit is,
\begin{equation}
    \tau_{\mathrm{damping}}^{\mathrm{rest}} = 49.80_{-7.49}^{+6.58} \left( \frac{M}{10^8 M_{\odot}} \right)^{0.43_{-0.04}^{+0.04}}
\end{equation}

with an intrinsic scatter of 0.11 ± 0.043 and the result of the Pearson correlation coefficient is r = 0.80. %The fitted plot can be seen in Figure \ref{fig:timescale vs M}.

In our PSD study, we see that in one of the blazar, the break time scale is consistent with the viscous time scale in the accretion disk suggesting a possible disk-jet coupling. In addition, the DRW results show that the non-thermal damping timescale produced in the jet for all three sources follows the same relation as the thermal damping time scale originating in the accretion disc. Both results support a possible disc-jet coupling scenario where it is believed that the long-term $\gamma$-ray variability could be caused by the fluctuation produced in the accretion disc which propagates to the jet.
%proving a disc-jet connection.
However, we present our next test in Section 3.4 where we produced the flux distribution for all three sources and discussed it in detail in the context of disc-jet connection.

%%%%%%%%%%%%%%   DRW modeling results
\begin{table*}
\centering
\setlength{\extrarowheight}{9pt}
\begin{tabular}{ccccc}
\hline
\hline
\multirow{2}{*}{\large{Sources}}  & \multicolumn{2}{c}{\large{Parameter of DRW}} & \multirow{2}{*}{$f^{\rm DRW}_{\rm b} (\rm day^{-1})$} & \multirow{2}{*}{$\tau^{\rm rest}_{\rm DRW}$ (day)}  \\
\cline{2-3}
& Log$\Large{\sigma}_{\mathrm{DRW}}$ & Log$\tau_{\mathrm{DRW}}$ \\
(1) & (2) & (3) & (4) & (5) \\
\hline

S4 0954+65  & $-0.79_{-0.09}^{+0.09}$ & $3.59_{-0.20}^{+0.21}$ & $0.0043_{-8.9\times 10^{-4}}^{+9.1\times 10^{-4}}$ & $265_{-48}^{+62}$ \\

PKS 0903-57  & $0.11_{-0.08}^{+0.08}$ & $3.59_{-0.15}^{+0.16}$ & $0.0043_{-6.6\times 10^{-4}}^{+7.0\times 10^{-4}}$ & $213_{-27}^{+39}$ \\

4C +01.02 & $0.10_{-0.13}^{+0.18}$ & $5.39_{-0.28}^{+0.37}$ & $0.0007_{-2.2\times 10^{-4}}^{+2.3\times 10^{-4}}$ & $707_{-172}^{+316}$ \\[+7pt]
\hline
\end{tabular}
\caption{\label{tab: Posterior parameters} The best-fit parameters of the DRW modeling are shown here. (1) source name, (2)-(3) posterior parameters of the DRW modeling of lightcurves, (4) break frequencies from DRW modeling, and (5) damping timescale in the rest frame.}
\end{table*}
%%%f^{\rm DRW}_{\rm b} (\rm day^{-1})$
%%% \tau^{\rm rest}_{\rm DRW}$ (days)
%%% S4 T_rest -  265_{-48}^{+62}
%%% PKS T_rest -  213_{-27}^{+39}
%%% 4C T_rest -  707_{-172}^{+316}

%%%%%%%%%%%%%%%%   M_BH and Damping timscale relation

%BHmass_rest_timescale_Burkefit.pdf
%BHmass_rest_timescale_Burke_Zhang_ourdata.pdf

\subsection{Flux distribution study}
 The flux distribution study helps us to understand the nature and origin of variability in the flux. %\sout{Log-normal flux distribution is observed in many blazars (\citealt{10.1093/mnrasl/sly136}, \citealt{10.1093/mnras/stz3108}) suggesting its multiplicative nature \citep{shah2018log} and generally can be produced in the disc or in the jet.} 
 A log-normal flux distribution has been observed in numerous blazars, manifesting at various wavelengths and timescales {\bf(\citealt{2010A&A...520A..83H}, \citealt{kushwaha2017gamma}, \citealt{shah2018log}, \citealt{10.1093/mnrasl/sly136}, \citealt{10.1093/mnras/stz3108}).} In the case of AGN and X-ray binaries, it is well established that the multiplicative process in the disc produces the log-normal flux distribution \citep{Uttley_2001}. If it is believed that the disc and jet in the blazar are somehow connected then the fluctuations produced in the disc can travel to the jet and produce the disc-induced variability imprinting the log-normal flux distribution. In \citet{Biteau2012}, they argue that the log-normal flux distribution can be produced by the minijets-in-a-jet statistical model, where the total flux is combined flux from many mini-jets which led to a skewed distribution. Their study also demonstrated that multiplicative processes can account for the log-normal flux distribution. Nevertheless, as indicated by \cite{Scargle2020}, the presence of a log-normal distribution in observed flux values does not necessarily imply the involvement of a multiplicative process.

%The origin of blazar variability across the wavebands is a long-standing question and very poorly understood. The variability can be identified in terms of flares, an occasional event when the flux rises above a certain level.
%However, in literature, many models have been proposed and one such model is the shock in the jet model which provides a good explanation for most of the objects. In the recent past, the minute scale variability observed in a few of the sources pointed toward a magnetic reconnection. Even considering these two models we do not clearly know if these variations are intrinsically produced inside the jet or if it has some connection with the accretion disk as well. The flux distribution study helps us to explore this possibility. 

% \usepackage{multirow}
\begin{table*}
\setlength{\extrarowheight}{3pt}
\begin{tabular}{|c|cc|cc|}
\hline
\multirow{2}{*}{Source} & \multicolumn{2}{c|}{Flux (Normal)}                        & \multicolumn{2}{c|}{Flux (Log-normal)}                      \\ \cline{2-5} 
                        & \multicolumn{1}{c|}{AD statistic} & p-value               & \multicolumn{1}{c|}{AD statistic} & p-value                \\ \hline
4C +01.02               & \multicolumn{1}{c|}{54.808}       & $< 2.2 \times 10^{-16}$& \multicolumn{1}{c|}{3.848}        & $1.34 \times 10^{-9}$\\ \hline
PKS 0903-57             & \multicolumn{1}{c|}{69.792}       & $< 2.2 \times 10^{-16}$& \multicolumn{1}{c|}{10.671}        & $< 2.2 \times 10^{-16}$\\ \hline
S4 0954+65              & \multicolumn{1}{c|}{39.281}       & $< 2.2 \times 10^{-16}$& \multicolumn{1}{c|}{4.8}        & $6.558 \times 10^{-12}$\\ \hline
\end{tabular}
\caption{\label{tab:ad-test} The AD statistic and corresponding p-values for the three sources. The null hypothesis of normal distribution is tested for flux and log-flux distributions.}
\end{table*}

\subsubsection{\bf Anderson-Darling test}
%Given a data sample the Anderson-Darling test, which is a modified version of the Kolmogorov-Smirnov test, is a statistical test to get the p-value for the data assuming that the data follows the Gaussian distribution. The Anderson-Darling test gives greater weight to the tails of the distribution compared to the Kolmogorov-Smirnov test. -->

 %The flux and log-flux data are tested for the null hypothesis of normal distribution. %From the FermiPy analysis, a test statistic (TS) value is generated for each data point along with the flux for the $\gamma$-ray light curve. 
 %The data points with TS $>$ 9 were selected for this analysis. We further filtered the data using the criteria $f_i / \sigma_i > 2$, where $f_i$ and $\sigma_i$ are the flux and the corresponding error for the ith data point. The results of the AD test are shown in Table \ref{tab:ad-test}. For a significance level of 0.05, we can reject the null hypotheses for the three sources 4C +01.02, PKS 0903-57, and S4 0954+65 since the p-values are less than 0.05. This means that both the normal and log-normal distributions are not a good fit for the flux data for the three sources. However, we note that the AD statistic values for the log-normal case are an order of magnitude smaller compared to the normal case. We explore this further in the following section.
 
 The Anderson-Darling (AD) test is a statistical method for evaluating the null hypothesis concerning the data following a Gaussian distribution. The null hypothesis probability value, p-value holds that when p-value > 0.05, it supports the data would indicate Gaussian distribution. Conversely, if the p-value is below 0.05, it indicates a departure from the Gaussian distribution.

In order to achieve the light curves with good statistics, the flux points with a TS value greater than 9 (i.e. TS $>$ 9) were selected.  Additionally, we refined the light curves by applying the condition that the ratio of the flux to its corresponding error ($f_i / \sigma_i > 2$) must exceed 2 for each individual data point. The results of the AD test for both linear-scaled and logarithmic-scaled flux data are shown in Table \ref{tab:ad-test}. The outcomes of the test statistic suggest that the flux distribution for the sources 4C +01.02, PKS 0903-57, and S4 0954+65 does not support either a normal or lognormal distribution. We note that the AD statistic values for the log-normal case are an order of magnitude smaller compared to the normal case. We explore this further in the following section.

\subsubsection{\bf Log-flux histogram}

\begin{comment}
\begin{figure*}
    \centering
    \includegraphics[width=\textwidth]{figures/histograms_with_errors.pdf}
    \caption{Histogram plots of flux, log-flux and spectral index for the sources 4C +01.02, PKS 0903-57, and S4 0954+65. The bins are generated using the Freedman-Diaconis (FD) rule.}
    \label{fig:histogram_flux_index}
\end{figure*}
\end{comment}

\begin{figure*}
\centering
\includegraphics[width=.33\linewidth]{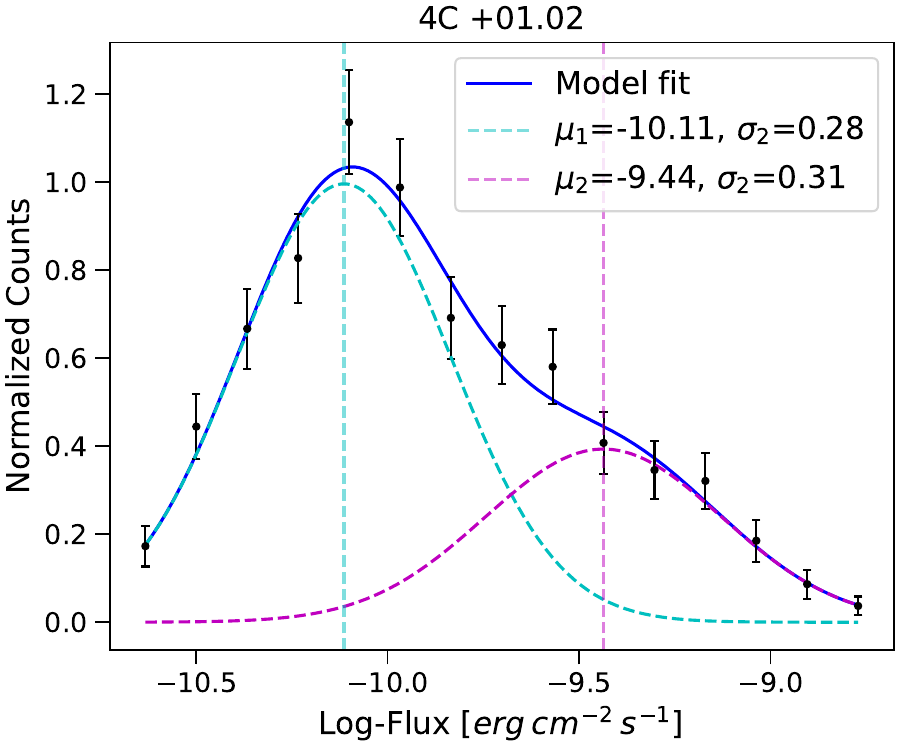}\hfill
\includegraphics[width=.33\linewidth]{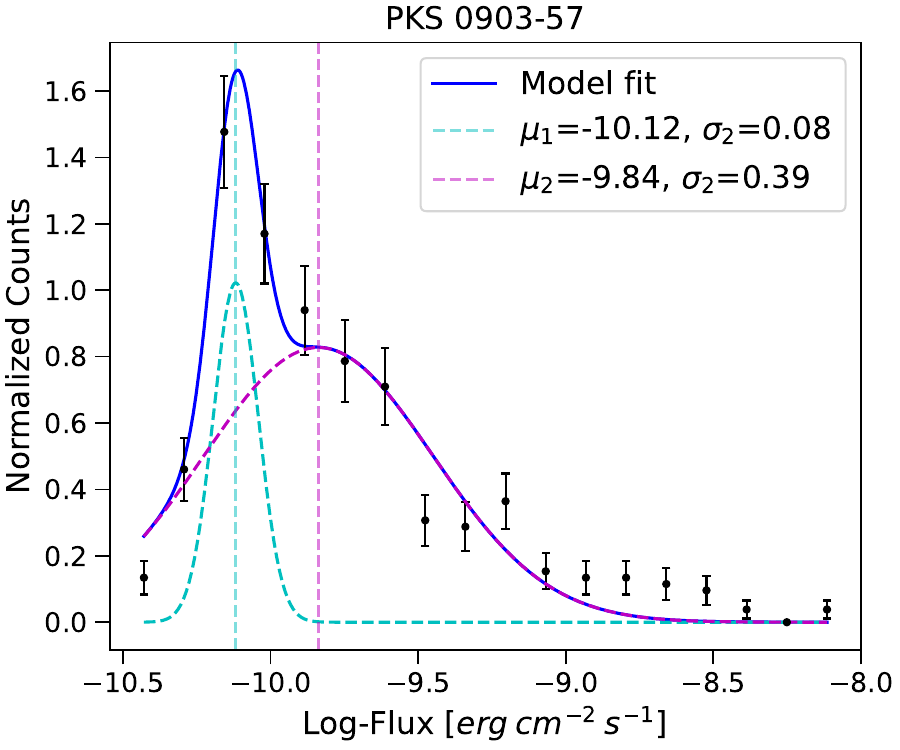}
\includegraphics[width=.33\linewidth]{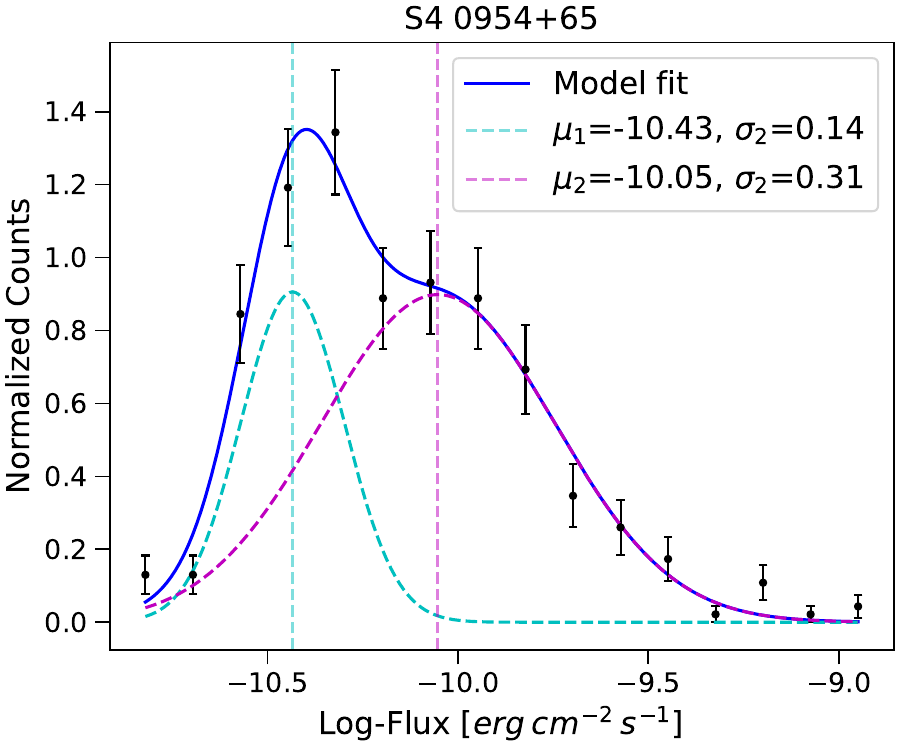}\vfill
\caption{The figure shows double log-normal fits of the $\gamma$-ray flux histogram for the sources 4C +01.02, PKS 0903-57, and S4 0954+65 (left to right). For each plot above, the solid blue line represents the double log-normal fit. The two scaled components with factors $a$ and $(1-a)$ of equation \ref{eqn:double_lognormal} are shown as cyan and magenta dashed curves respectively. The vertical dashed lines are at $\mu_1$ and $\mu_2$ positions. The corresponding parameters for fit are shown in table \ref{tab:double_lognormal}.}

\label{fig:double_lognormal}
\end{figure*}
%Histograms of the flux and log of flux can give more insight into their distributions. 
%We analyze the log-flux histogram in this section. %It has a more centered distribution while the flux histogram is right-skewed. 
%The data is binned using the Freedman-Diaconis (FD) rule. Further, for each bin, the bin width is greater than the average flux error of the data points in the bin \citep{shah2018log}. The histogram is normalized and it shows a double-peaked structure. 
We generate the normalized histogram of the logarithmic flux values for the three selected sources. The flux values are grouped into bins using the Freedman-Diaconis (FD) rule. The bin width of the histogram is decided such that it is greater than the average flux error of the data points within that bin \citep{shah2018log}. The flux histogram for the sources shows a double-peaked structure. Therefore, the log-flux histogram is fitted using the double Gaussian probability density functions (PDF) given by
\begin{equation}
    f(x) = 
    \frac{a}{\sqrt{2 \pi} \sigma_1} \exp{\frac{-(x-\mu_1)^2}{2 \sigma_1^2}} + 
    \frac{(1-a)}{\sqrt{2 \pi} \sigma_2} \exp{\frac{-(x-\mu_2)^2}{2 \sigma_2^2}}
    \label{eqn:double_lognormal}
\end{equation}
where $\mu_1$ and $\mu_2$ are mean values of the two Gaussian functions and $\sigma_1$ and $\sigma_2$ are their corresponding standard deviations. The functions are multiplied by scaling factors $a$ and $1-a$. The fits are shown in Figure \ref{fig:double_lognormal} and the best-fit parameters are reported in Table \ref{tab:double_lognormal}. The reduced-$\chi^2$ statistic is used to estimate the quality of the fit.

%\textcolor{magenta}{When the distribution is normal with respect to log-flux it is log-normal with respect to flux. -> no need to mention, just for your understanding: lognormal flux distribution is when the flux distribution is Gaussian on the logarithmic scale}So the equation \ref{eqn:double_lognormal} is double log-normal with respect to the flux when it is applied to the log-flux distribution. 
%Apart from the above equation, we also tried other combinations of distributions such as double normal, single normal, and double log-normal. However, the reduced-$\chi^2$ values were less optimal compared to the double log-normal fits presented in table \ref{tab:double_lognormal} 
Furthermore, we investigate the double-peaked histogram with the other combinations of distribution functions such as double Gaussian, a combination of single Gaussian and log-normal, and double log-normal. However, the obtained reduced-$\chi^2$ values exhibited less favorable results as compared to the double log-normal fits presented in Table \ref{tab:double_lognormal} . The vertical lines in Figure \ref{fig:double_lognormal} representing $\mu_1$ and $\mu_2$ are closer in the case of PKS 0903-57 (middle panel) compared to the other two sources and the standard deviation parameter $\sigma_2$ is much higher than $\sigma_1$ which is consistent with the data. The scaling factor $a$ is 0.20 which means the right component (dashed magenta curve) is dominant resulting in a lower profile of the left component (dashed cyan curve). The case of S4 0954+65 is similar. However, for 4C +01.02 the left component is dominant with a value 0.70. As we mentioned earlier for flux distribution study we choose the data points which have TS $>$ 9 which by default removes the upper limits. We tested the flux distribution by including the upper limits as well but it did not alter the existing shape of the distribution.

\begin{table*}
\setlength{\extrarowheight}{3pt}
\begin{tabular}{|c|c|c|c|c|c|c|c|c|}
\hline
Source      & $\mu_1$ & $\mu_2$ & $\sigma_1$ & $\sigma_2$ & $a$ & DoF & $\chi^2 / DoF$ \\ \hline
4C +01.02   & -10.11   & -9.44   & 0.28      & 0.31       & 0.70  & 14  & 0.57         \\ \hline
PKS 0903-57 & -10.12   & -9.84   & 0.08      & 0.39       & 0.20  & 17  & 3.46         \\ \hline
S4 0954+65  & -10.43   & -10.05   & 0.14      & 0.31       & 0.31  & 15  & 1.86         \\ \hline
\end{tabular}
\caption{\label{tab:double_lognormal} Results of the double Gaussian fit on the log-flux data for the three sources.}
\end{table*}

\section{DISCUSSION}
We present a comprehensive investigation aimed at characterizing the $\gamma$-ray variability exhibited by some selected Fermi blazars. Employing various approaches and methodologies, we analyze the light curves to elucidate the essence of $\gamma$-ray emissions and to understand the interplay between the accretion disc and relativistic jets.

From the analysis of fractional variability, as illustrated in Figure \ref{fig:Fvar vs energy} and summarized in Table \ref{tab: Fvar vs energy}, for S4 0954+65, the UV and $\gamma$-ray emissions exhibit higher variability compared to its optical and X-ray emissions. In addition, a dip in the X-ray band resulted in a double-hump feature in the Fvar distribution. In the cases of PKS 0903-57 and 4C +01.02, the $\gamma$-ray emissions display a higher level of variability in comparison with the optical/UV and X-ray emissions. Specifically, for PKS 0903-57, an increment in $\gamma$-ray variability is approximately fourfold when compared to the UV band (see middle panel of Figure \ref{fig:Fvar vs energy}), while for 4C +01.02, it increases by a factor of $\sim$2.5 (see right panel of Figure \ref{fig:Fvar vs energy}).

A similar variability pattern has been reported in other blazars \citet{Abdo_2011, Rani2016, shah2021unveiling, malik2022multiwavelength, Prince23} as well. In the case of S4 0954+65, the observed distinctive pattern in Fvar distribution suggests that optical/UV and $\gamma$-ray emissions are associated with the synchrotron and inverse Compton (IC) processes involving high-energy electrons. On the other hand, the X-ray emission is linked to the IC process involving the lower-energy tail of the electron distribution. This observed variability can be attributed to relativistic electron cooling, wherein higher-energy electrons undergo more rapid cooling compared to their lower-energy counterparts. As a result, we observe higher variability in the emissions originating from high-energy electrons.

%%%%%%%%%%%%%%%%%%%%%%%%%%%%%%%%%%%%%%%%%%%%%%%%%%    
 %In the context of the depiction of periodograms, 
 The PSD in AGN is well fitted with a single power law of  form $P(\nu) \propto \nu^{-\beta}$, where the slope value spans the interval from 0 to 2 \citep{sobolewska2014stochastic}. A case study focusing on a sample of blazars with $\gamma$-ray observations was conducted by \citep{bhatta2020nature}. The results of their investigation revealed that the slope index spans a range from 0.8 to 1.5. When $\beta$ equals 1, it is commonly referred to as long-term flicker noise, which can appear coherent for extended timescales (spanning decades). Our investigation involves fitting PSDs for all three sources using both single power law and bending power-law models. The obtained index values for the single power-law model are 0.80, 1.28, and 1.32 for S40954+65, PKS0903-57, and 4C+01.02, respectively. In addition, we observed breaks in frequency at 0.0047 (day$^{-1}$) for S4 0954+65, 0.004 (day$^{-1}$) for PKS 0903-57, and 0.0014 (day$^{-1}$) for 4C +01.02. The frequency break in PSD deduced from the DRW modeling is consistent within the error bar with the PSD break obtained from the observed data. In the case of 4C +01.02, the estimated PSD break from the data is not consistent with the DRW modeling suggesting the length of the light curve is not sufficient to derive a true PSD shape. Notably, the observed emissions are Doppler-boosted from the jets. The observed break frequency could potentially signify the impact of accretion disc events on the jets, driven through various mechanisms like accretion rates, magnetic fields, viscosity, and disc instabilities. These dynamic events might serve as drivers of jet emission variability. We compare the break time scale of PKS 0903-57 with the viscous time scale in the accretion disc and the timescale was found to be consistent suggesting a possible disc-jet coupling in the source. \par

 %%%%%%%%%%%%%%%%%%%%%%%%%%%%%%%%%%%%%%
 %%%%%%%%  In the context of DRW modeling
 Several authors have carried out an extensive investigation to explore the characteristic timescale of optical variability of AGN accretion disc \citep{collier2001characteristic, kelly2009variations, macleod2010modeling, simm2016pan, suberlak2021improving}. Additionally, \citet{ryan2019characteristic, zhang2022characterizing} extended these methodologies to investigate $\gamma$-ray variability within blazar samples. %\citet{zhang2022characterizing} show  that the characteristic rest frame timescales of $\gamma$-ray emission are consistent with the optical time scale derived for AGN in \citet{Burke_2021}. 
 %with host black hole mass values, superimposing these findings upon the outcomes of \citet{Burke_2021} and observed a convergence that the $\gamma$-ray variability timescale aligns with the optical timescale, indicating a link to shared underlying physical mechanisms in accretion disk phenomena in different astrophysical objects. In the context of AGN that directly affects the emission from the jet.

%\citet{Yang_2021} model the 11 yrs of $\gamma$-ray light curve of sample of blazar from 4FGL catalog. They aimed to explain the $\gamma$-ray QPO time scale by modeling the light curve with Gaussian and celerite processes. Their results suggest that the $\gamma$-ray variability is a Gaussian process. 

\citet{zhang2022characterizing} used the $\sim$12.7 years of $\gamma$-ray light curve of a sample of 23 AGN and modeled the variability with the stochastically driven damped simple harmonic oscillator (SHO) and the damped random-walk (DRW) models. In most of their cases, they find DRW explains the variability better than the SHO model. From the DRW modeling, they derived the damped random time scale which corresponds to the frequency break in the PSD, and plot it against the black hole mass. Their $\gamma$-ray sources almost occupy the same space in the $\tau^{rest}_{damping}$ -- M$_{BH}$ plot as the optical light curve derived for AGN in \citet{Burke_2021}, and they conclude that the observed break or DRW time scale in $\gamma$-ray corresponds to the processes in the accretion disc, suggesting a possible disc-jet coupling. Later, \citet{Zhang_2023} used the broadband light curve from radio to $\gamma$-ray and modeled the variability with the DRW process. Their results suggest that the non-thermal optical, X-ray, and $\gamma$-ray are produced at the same location and radio at a farther distance. The derived time scale for non-thermal optical and $\gamma$-ray emission occupies the same space as thermal emission from the disc, suggesting the possible disc-jet coupling.

The derived damped random time scale for our three sources is also plotted on $\tau^{rest}_{damping}$ -- M$_{BH}$ space along with thermal time scale from \citet{Burke_2021} and non-thermal time scale from \citet{Zhang_2023}. The plots are shown in Figure~\ref{fig:timescale vs M}. First, we plot our three sources along with the disc time scale derived for AGN from \citet{Burke_2021}, we noticed that our sources exactly lying on the best-fit relation derived by \citet{Burke_2021}, suggesting the non-thermal $\gamma$-ray time scale in our three sources matching with thermal optical time scale derived for AGN. This led to the conclusion that the variability in the $\gamma$-ray light curve is linked with disc variability proving a possible disc-jet coupling. The non-thermal time scale derived for our three sources are also matching with the non-thermal time scale from \citet{zhang2022characterizing} as shown in Figure~\ref{fig:timescale vs M}. We fitted all the sources together to derive a combined standard relation for $\tau^{rest}_{damping}$ -- M$_{BH}$ and the slope is consistent within the error bar with \citet{Burke_2021} and \citet{zhang2022characterizing}.

%%%%%%%%%%%%%%%%%%%%%%%%%%%%%%%
%%%%%%%%%%%%   Flux distribution
To perform the $\gamma$-ray flux distribution study, we use weekly binned Fermi-LAT light curve. The light curve was analyzed by the Anderson-Darling test. The test for normality of both the flux and log-flux was rejected due to the low p-values. The normalized log-flux distribution was further analyzed by fitting a double normal distribution function (see equation \ref{eqn:double_lognormal}) owing to its double-peaked structure histogram, suggesting the presence of two distinct flux states. The fits for sources 4C +01.02 and PKS 0903-57 have $reduced-\chi^2$ values close to 1. %%{Other combinations of normal and log-normal double distributions were tried but we found that the double normal (double log-normal) distribution with respect to log-flux (flux) was optimal for all three sources even though the $reduced-\chi^2$ value for S4 0954+65 is 1.6 (see table \ref{tab:double_lognormal})}. 
 Alternative combinations of normal and log-normal distribution functions were also investigated. However, it was observed that the most suitable choice for all three sources remained the double log-normal distribution. This conclusion holds even though the $reduced-\chi^2$ values range from 0.57 to 3.46 (see table \ref{tab:double_lognormal}).\par
As the X-ray binaries study suggests the variability produced in the accretion disc is caused by multiplicative processes which led to a log-normal flux distribution \citep{Uttley_2001, Uttley_2005}. 
Our study shows that there is some connection between variability produced in the jet with the accretion disc. Therefore, we also produced the flux distribution of the $\gamma$-ray light curve for our sources. 
To mitigate bias in the flux distribution resulting from the possible lack of lower luminosity flux states, we have removed the flux with larger uncertainty, where $f_i / \sigma_i < 2$. 
We noticed that the distribution rather follows a double log-normal flux distribution. %\sout{A double log-normal flux distribution across the wavebands is also seen in PKS 1510-089 \citep{kushwaha2017gamma}}.
In the investigation of multi-wavelength flux distribution in PKS 1510-089, \cite{kushwaha2017gamma} discovered two distinct log-normal profiles within the flux distribution at near-infrared (NIR), optical, and ${\gamma}$-ray wavelengths, which are associated with two possible flux conditions in the source. In a separate study by \cite{10.1093/mnras/stz3108}, a dual Gaussian flux distribution was observed in the X-ray energy band for Mkn 501 and Mkn 421, providing additional support for the hypothesis of two distinct flux states.
The origin of the double log-normal distribution is still unknown and the study done by \citet{Stecker_1996, Stecker_2011} and \citet{kushwaha2017gamma} suggests it could be coming from the extragalactic $\gamma$-ray background.

\section{Conclusion}
In this study, we delved into several facets of multi-waveband variability exhibited by three distinct blazars, employing a range of methodologies. We emphasized our interest in characterizing the $\gamma$-ray variability. The findings of our investigation lead us to the following conclusions: 
\begin{itemize}
    \item %%%%%%%{We performed a comprehensive study involving fractional variability analysis using broadband light curves for three sources. Notably, the observed multi-waveband variability pattern resembles the typical double-hump profile at least in one of the sources as we generally see in the broadband SED, a hallmark feature crucial for discerning the emission mechanism nature inherent to blazars.\par}
    We performed a comprehensive study involving fractional variability analysis using broadband light curves for three sources. Notably, in one of the sources, the observed variability pattern resembles a double-hump profile, as generally seen in broadband spectral energy distribution (SED) in blazars.
    \item To comprehensively study the statistical characteristics of the variability within $\gamma$-ray light curves over a wide range of temporal frequencies, we performed PSD analysis with extensive MC simulations. This study revealed a break in the $\gamma$-ray PSD for two sources. These observed break frequencies were subsequently confirmed through DRW modeling applied to $\gamma$-ray light curves. 
    The observed PSD break time scale for PKS 0903-57 is consistent with the estimated viscous time scale in the accretion disk suggesting a possible scenario of disk-jet coupling.
   %for the PKS 0903-57 source, an intriguing observation emerged: the detected variability timescale is consistent with the viscous timescale in the accretion disk. This observation suggests the possibility of a disk-jet coupling within this particular source.
    \item %\sout{Examining the flux distribution, the manifestation of a log-normal distribution in blazar flux could potentially signify the involvement of multiplicative processes. These processes might entail the interplay of perturbations occurring within both the disk and/or the jet. {\bf The observed flux distribution in all three sources is well described by a bi-model log-normal distribution. }}
    Analyzing the flux distribution, the presence of a log-normal distribution in blazar flux might indicate the involvement of multiplicative processes. These processes could involve perturbations in both the disk and/or the jet. The observed flux distribution in all three sources is accurately described by a double log-normal distribution, which has also been identified in earlier studies.
    %However, our investigation led to a remarkable revelation: the flux distribution is well described by a dual log-normal distribution. 
    This finding points to a complex scenario where underlying flaring activity emerges as a result of the combined variability of long and short timescales.
    \item The non-thermal damping time scale derived from the DRW modeling is consistent with the thermal damping time scale derived for optical AGN suggesting a possible disk-jet coupling in these three blazars. This serves as a motivation for studying a larger sample including broadband light curves to derive a fair conclusion.
   % However, this study needs to be extended over a larger sample including broadband light curves to derive a fair conclusion.
    
\end{itemize}
\section*{Acknowledgements}
  We thank the referee for their insightful comments and suggestions.
  A. Sharma is grateful to Prof. Sakuntala Chatterjee at S. N. Bose National Centre for Basic Sciences, for providing the necessary support to conduct this research.
  R. Prince is grateful for the support from Polish Funding Agency National Science Centre, project 2017/26/A/ST9/-00756 (MAESTRO 9) and the European Research Council (ERC) under the European Union’s Horizon 2020 research and innovation program (grant agreement No. [951549]). 

%%%%%%%%%%%%%%%%%%%%%%%%%%%%%%%%%%%%%%%%%%%%%%%%%%
\section*{Data Availability}

This research has used $\gamma$-ray observations from Fermi-LAT and it can be accessed from publicly available LAT database server at \url{https://fermi.gsfc.nasa.gov/ssc/data/access/}. Additionally, this study also used data available in the X-ray and UVOT wavebands from the Neil Gehrels Swift Observatory and can be accessed from \url{https://heasarc.gsfc.nasa.gov/w3browse/swift/swiftmastr.html}.



\bibliographystyle{mnras}
\bibliography{bibliography} % if your bibtex file is called example.bib

\bibliographystyle{mnras}
%\bibliography{bibliography} 

% Don't change these lines
\bsp	% typesetting comment
\label{lastpage}
\end{document}